\documentclass[aps,pra,twocolumn,,superscriptaddress,floatfix]{revtex4-2}
\usepackage[svgnames]{xcolor}
\definecolor{darkred}{rgb}{0.6, 0, 0}
\definecolor{darkgreen}{rgb}{0, 0.5, 0}
\usepackage[colorlinks=true,urlcolor = orange,linkcolor = darkgreen, citecolor = darkred, bookmarks=false]{hyperref}
\usepackage{amsfonts,amsmath,amssymb,amsthm}
\usepackage{graphicx}
\usepackage{dsfont}
\usepackage{cleveref}
\usepackage[normalem]{ulem}
\usepackage{enumerate}
\usepackage{natbib}
\usepackage{diagbox}
\usepackage{enumitem}
\usepackage{mathrsfs}
\usepackage{esint}

\usepackage{tcolorbox}
\usepackage{tabularx}
\usepackage{array}
\usepackage{colortbl}
\tcbuselibrary{skins}

\swapnumbers

\usepackage{tcolorbox}
\usepackage{tabularx}
\usepackage{array}
\usepackage{colortbl}
\tcbuselibrary{skins}
\newcolumntype{C}{>{\centering\arraybackslash}X}
\newcolumntype{Y}{>{\raggedleft\arraybackslash}X}

\def\ii{{\rm i}}
\newcommand{\dd}{{\rm d}}

\usepackage{tikz}
\usepackage{pgfplots}
\usetikzlibrary{arrows,topaths,shapes.geometric,shapes.misc,patterns,positioning,shadows,decorations.markings,
decorations.pathreplacing,calc, trees, positioning, arrows, shapes, shapes.multipart, shadows, matrix, decorations.pathreplacing, decorations.pathmorphing}
\tikzset{->-/.style={decoration={
  markings,
  mark=at position #1 with {\arrow[scale=1.5]{latex}}},postaction={decorate}}}

\usepackage{pgfplots}
\pgfplotsset{compat=1.9}

\begin{document}

\title{Single-file dynamics with general charge measures}

\author{\v{Z}iga Krajnik}
\affiliation{Department of Physics, NYU, 726 Broadway, New York, NY 10003, United States}

\date{\today}

\begin{abstract}
We study charge fluctuations in single-file dynamics with general charge measures. The exact finite-time distribution of charge fluctuations is obtained in terms of a dressing transformation acting on the finite-time distribution of particle fluctuations. The transformation is mapped to a simple substitution rule for corresponding full-counting statistics. By taking the asymptotics of the dressing transformation, we analyze typical and large scale charge fluctuations. Typical charge fluctuations in equilibrium states with vanishing mean charge are anomalous, while large charge fluctuations undergo first and second order dynamical phase transitions out of equilibrium.
\end{abstract}

\maketitle

\section{Introduction}\label{sec1}
One of the most widely applicable results of probability theory is the central limit theorem that, in its basic form, characterizes fluctuations of the average of $n$ independent identically distributed variables sampled from a distribution with finite variance. The theorem states that typical fluctuations around the mean value are of order $\mathcal{O}(n^{1/2})$ and their distribution is asymptotically Gaussian. The central limit theorem has been further refined to cover the important cases of not identically distributed or weakly correlated variables \cite{billingsley1995probability_book}.

A natural extension to the study of typical fluctuations is given by large deviation theory, characterizing large fluctuations of order $\mathcal{O}(n)$ away from the mean. A basic problem of large deviation theory is to establish a large deviation principle for tail probabilities, i.e. to show that large fluctuations are suppressed exponentially in $n$ and to characterize the rate of suppression, encapsulated in the rate function. The central limit theorem can then often be understood as the quadratic expansion of the rate function around the mean value. 

The theory of large deviations is also the natural language to describe the central notion of concentration of measure in statistical mechanics both in and out of equilibrium \cite{Oono_1989,Touchette_LDT}. 
Fluctuations of conserved quantities in many-body systems are captured by the full-counting statistic of corresponding time-integrated current densities,
the time-integrated current being the many-body analogue of the sum of a sequence of variables, with the proviso that current densities at different times are now correlated by the dynamics with randomness necessarily entering even for deterministic dynamics due to sampling from an ensemble of initial conditions.

This naturally leads to the problem of characterizing current fluctuations on both the large and typical scale. The question of large fluctuations in ergodic systems with a single conserved quantity has been thoroughly addressed by the development of macroscopic fluctuations theory, see \cite{MFT} for a review. The theory establishes a large deviation principle for fluctuations of the integrated current and formulates a variational problem, the solution of which recovers the rate function. Recently, building upon the theory of generalized hydrodynamics \cite{GHD_Bertini,GHD_Doyon}, an analogous ballistic macroscopic fluctuations theory \cite{BMFT} has been developed to describe large fluctuations in integrable systems. Similar results have also been derived by mapping the full-counting statistics in a non-equilibrium bipartite ensemble to a dual equilibrium problem using space-time duality \cite{SpaceTimeDuality}.

Establishing a large deviation principle quantifies the structure of large fluctuations and its expansion to quadratic order suggest asymptotic Gaussianity of typical fluctuations. This might lead to the erroneous conclusion that a large deviation principle directly implies Gaussianity on the typical scale. However, inferring behavior on the typical scale from the large scale involves a rescaling for which the logarithm of the full-counting statistics needs to satisfy an additional regularity condition \cite{Bryc1993}. While Gaussianity of typical fluctuations in a class of chaotic quantum systems has been recently demonstrated  \cite{QuantumChaoticFCS}, numerical studies of classical integrable spin chains \cite{Krajnik2022a,Krajnik2024} have found clear violations of asymptotic Gaussianity in equilibrium ensembles at half-filling.

This unexpected observation has led to an exact solution of equilibrium fluctuations \cite{Krajnik2022} in a simple interacting classical cellular automaton of impenetrable ballistic $\mathbb{Z}_2$-charged particles first studied in \cite{Klobas2018}. Typical fluctuations were found to follow an M-Wright distribution, normally associated with fractional diffusion \cite{Mainardi1996,Mainardi_2020,PhysRevE.61.132,Barkai_2002}.
The violation of Gaussianity was traced to a complex zero of the full-counting statistics colliding with the origin, breaking the regularity condition in a process reminiscent of an equilibrium phase transition. Shortly thereafter, the same distribution was derived in the low temperature regime of the sine-Gordon model \cite{Kormos}, see also \cite{Altshuler_2006} for an earlier derivation in the same setting using form factors. It was soon realized that the observed phenomenology is closely related to fragmentation of the configuration space by a strict kinetic constraint, giving rise to the class of charged single-file dynamics whose fluctuations were studied in detail in \cite{Krajnik2022SF}. Besides anomalous fluctuations in equilibrium, the non-equilibrium setting gave rise to a rich interplay of dynamical phase transitions in the rate function of large fluctuations.
Apart from fluctuations, single-file kinetic constraints have also been used to study dynamical correlation functions at hydrodynamics scales \cite{Knap} and in some exact solutions \cite{SFBetheAnsatz,PhysRevLett.128.130603,Zadnik_Bocini_Bidzhiev_Fagotti_2022}.

Nevertheless, the class of dynamics studied in \cite{Krajnik2022SF} had two important limitations. Firstly, it was not clear what role was played by the specifics of the $\mathbb{Z}_2$ charge measure. Secondly, the dynamic were restricted to particles with inert charges, whose interaction came solely from their impenetrability and not the charges themselves. In the present work, we relax the first of these assumptions and show that the phenomenology of fluctuations established in \cite{Krajnik2022SF} remains intact on both the large and typical scale, requiring only mild restrictions on the charge measure. 

The paper is organized as follows: In Section \ref{sec:CSF}, we introduce single-file dynamics and the basic concepts used in the remainder of the work.
In Section \ref{sec:dressing} we derive the exact finite-time joint charge-particle fluctuations in terms of particle fluctuations.
In Section \ref{sec:asymptotics} we study the time-asymptotic behavior of joint fluctuations at both the typical and large scale.
We conclude in Section \ref{sec13} by putting our results in the context of recent advances in the study of fluctuations in integrable systems and related models.

\section{Charged single-file dynamics}
\label{sec:CSF}
Consider a system of $Q_p$ classical particles on a periodic ring of length $L$ and label particle positions in configuration space $\mathcal{Q}_p \equiv [-L/2, L/2)^{Q_p}$ at time $t$ as ${\bf x}(t) \equiv (x_1(t), \ldots, x_{Q_p}(t)) \in \mathcal{Q}_p$, where we identify $x_j = x_j + L$. Assuming a generic position (one in which no two particles coincide),
we order the particles from left to right at initial time $t_0$,
$x_1(t_0) < x_2(t_0) < \ldots < x_{Q_p}(t_0).$
The particles' dynamics is given by a map $\phi_{s}: \mathcal{Q}_p \to \mathcal{Q}_p$ which propagates the particles' positions forward in time
\begin{equation}
	\phi_s\left[{\bf x}(t)\right] = {\bf x}(t+s). \label{U_def}
\end{equation}
We consider dynamics $\phi_s$ that admit a flat invariant measure characterized by the average particle density $\rho$
and specialize to a simple nonequilibrium setting by considering a bipartite initial measure consisting of two equilibrium measures joined at the origin with particle densities $\rho_\pm$ to the right/left of the origin respectively
\begin{equation}
	\varrho_L({\bf x}, t_0) = \prod_{j=1}^{Q_p}p_L(x_j),  \label{neq_def} 
\end{equation}
with the one particle density
\begin{equation}
	p_L(x) =
	\frac{2L^{-1}}{\rho_- + \rho_+}
	\begin{cases}
		\rho_-, & -L/2 \leq x < 0,\\
		\rho_+, & 0 \leq x < L/2.
	\end{cases}
\end{equation}
The integral of the local particle density $\rho_p(x, t) \equiv \sum_{n=1}^{Q_p} \delta(x-x_n(t))$ gives the total particle number $Q_p = \int_{-L/2}^{L/2}\dd x \, \rho_p(x, t)$.  We consider systems in the thermodynamics limit at finite particle density with the corresponding initial measure
\begin{equation}
	\varrho({\bf x}, t_0) = \lim_{\stackrel{Q_p, L \to \infty}{Q_p/L = (\rho_-+\rho_+)/2}} \varrho_L({\bf x},t_0). \label{neq_def_td}
\end{equation}

\begin{figure}[h!]
	\includegraphics[width=\linewidth]{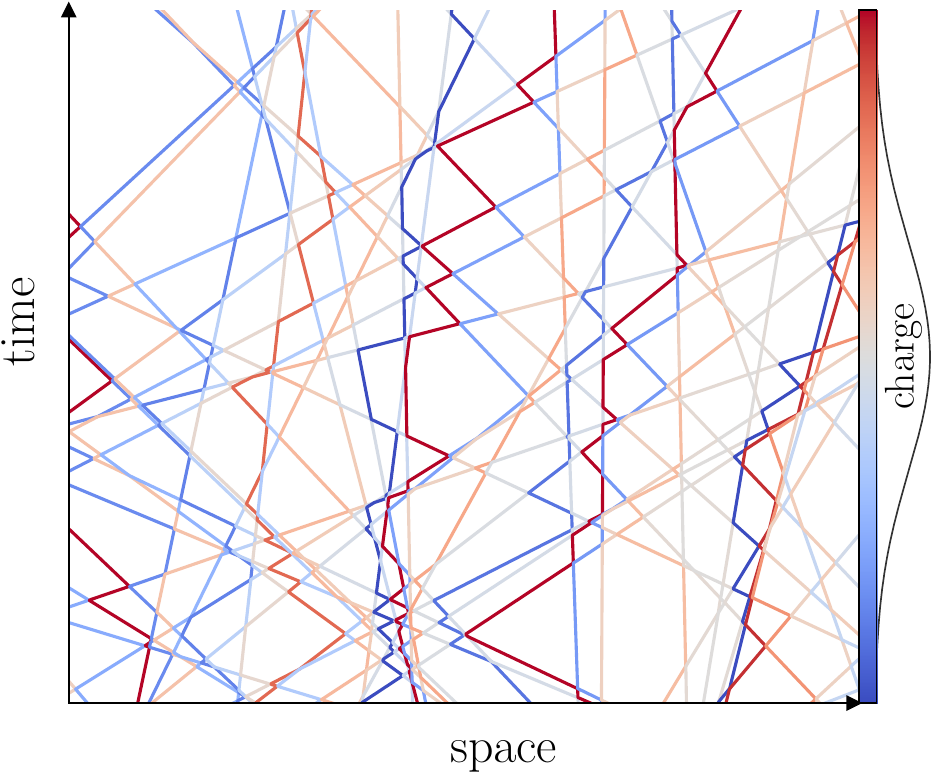}
	\caption{Many-body trajectory of an inert charged single-file system consisting of ballistically propagating hardcore particles (broken straight lines) carrying inert charges (blue to red) sampled from a Gaussian charge measure \eqref{Gauss_measure}.}
	\label{fig:sf}
\end{figure}

We attach a scalar charge $c \in \mathcal{C} \subseteq \mathbb{R}$ to each particle, sampled from a normalized measure,
$\omega_\mu: \mathcal{C} \to \mathbb{R}^+_0$, $\int_{\mathcal{C}} \dd \omega_\mu = 1$, with a finite mean $|\int_{\mathcal{C}} \dd \omega_\mu\, c|  = |\mu| < \infty$ and variance $\int_\mathcal{C} \dd \omega_\mu (c-\mu)^2  = \sigma^2_\omega < \infty$ whose support is not sign-definite, ${\rm s}_\pm \equiv {\rm supp}(\omega_\mu) \cap \mathbb{R^\pm} \neq \emptyset$ and denote $\pm c_{\pm} \equiv \displaystyle  \max_{c \in {\rm s}_\pm} \pm c$.

The enlarged configuration space $\mathcal{Q}^{c}_p \equiv \mathcal{Q}_p \times \mathcal{Q}_c$ is spanned by  $({\bf x}(t), {\bf c})$, where ${\bf c} \equiv (c_1,\ldots,c_{Q_p}) \in \mathcal{Q}_c \equiv \mathcal{C}^{Q_p}$ and $c_j$ is the charge of the $j$-th particle.
We consider \emph{inert} charges, meaning that the dynamics of the charged system $\phi_s^{c}: \mathcal{Q}^{c}_p \to \mathcal{Q}^{c}_p$ decomposes as 
\begin{equation}
	\phi_s^{c}[\left({\bf x}(t), {\bf c}\right)] = \left(\phi_s[{\bf x}(t)], {\bf c}\right) = \left({\bf x}(t+s), {\bf c}\right). \label{Uc_def}
\end{equation}
The integrated local charge density $\rho_c(x, t) \equiv \sum_{i=1}^{Q_p} c_i \delta(x - x_i(t))$ gives the total charge $Q_c \equiv \int_{-L/2}^{L/2} \dd x\, \rho_c(x,t)$.
We extend the bipartite particle initial measure \eqref{neq_def} to $\mathcal{Q}_p^{c}$ by specifying the mean charge values $\mu_\pm$ in each partition
\begin{equation}
	\varrho^{c}_L({\bf x}, {\bf c}, t_0) = \prod_{j=1}^{Q_p}p_L(x_j)  \omega_{\mu_{s_j}}(c_j), \label{neq_comp_def}
\end{equation}
where $s_j=\textrm{sgn}(x_j)$ and take the thermodynamic limit
\begin{equation}
	\varrho^{c}({\bf x}, {\bf c}, t_0) = \lim_{\stackrel{Q_p, L \to \infty}{Q_p/L = 
			(\rho_-+\rho_+)/2}} \varrho^{c}_L({\bf x}, {\bf c},t_0). \label{neq_comp_def_td}
\end{equation}
We further consider \emph{single-file}  dynamics for which the initial particle ordering is preserved by the dynamics $\phi_s^{c}$
\begin{equation}
	x_1(t) < x_2(t) < \ldots < x_{N-1}(t) < x_N(t), \quad \forall t, \label{cond2}
\end{equation}
which, due to charge inertness \eqref{Uc_def}, is the same as imposing the single-file constraint on the particle dynamics $\phi_s$. A snapshot of an inert charged single-file system in the thermodynamic limit is shown in Figure \ref{fig:sf}.
Note that in an arbitrary crossing (i.e.~one violating the single-file condition \eqref{cond2}) particle dynamics it is always possible to enforce the single-file constraint by adding a strong repulsive hardcore interaction to the particles.
\subsection{Full-counting statistics}
Conservation of particle number $Q_p$ and total charge $Q_c$ means that their local densities satisfy  continuity equations
$\partial_t \rho_i(x, t) + \partial_x j_i(x, t) = 0$,
where $i \in \{p, c\}$ and $j_i(x, t)$ is the local current density. The corresponding integrated currents at the origin 
\begin{equation}
	J_{i}(t) \equiv \int_{t_0}^{t_0+t}\dd t'\, j_{i}(0, t'), \label{Ji_def}
\end{equation}
equal the total charge transported across the origin $J_i(t) = Q_i^+(t_0+t) - Q_i^+(t_0)$ in an interval of time $t$,
where $Q_i^+(t) \equiv \int_0^{\infty} \dd x\, \rho_i(x, t)$.
Fluctuations of transported charges are encoded in the joint full-counting statistics
\begin{equation}
	G_{c,p}(\lambda_c, \lambda_p|t) \equiv \langle e^{\lambda_c J_c(t) + \lambda_p J_p(t)} \rangle_{\mathcal{Q}_p^{c}}, \label{FCS_def}
\end{equation}
where $\lambda_{p}$ and $\lambda_c$ are the particle and charge counting fields respectively and the average $\langle \bullet \rangle_{\mathcal{Q}_p^{c}}$ is taken with respect to the initial nonequilibrium measure \eqref{neq_comp_def_td}. The corresponding joint probability distribution $\mathcal{P}_{c,p}(J_c, J_p|t)$ is recovered by the inverse Laplace transform of the full-counting statistics
\begin{equation}
	\mathcal{P}_{c, p}(J_c, J_p|t) = \mathcal{L}^{-2}_{\lambda_c, \lambda_p}\left[ G_{c, p}(-\lambda_c, -\lambda_p|t) \right](J_c, J_p), \label{Laplace_P_G}
\end{equation}
where $\mathcal{L}_{x}[f(x)](y) \equiv \int_{-\infty}^\infty \dd x\, e^{-y x} f(x)$ is the bilateral Laplace transform with respect to $x$. The univariate probability distributions $\mathcal{P}_i(J_i|t)$ of the integrated particle/charge current are obtained by marginalizing the joint probability distribution, which at the level of full-counting statistics  amounts to setting a counting field to zero, $G_i(\lambda_i|t) = G_{c, p}(\lambda_c, \lambda_p|t)|_{\lambda_{j\neq i}=0}$.
The charge full-counting statistics $G_c$ involves averaging over the composite initial measure \eqref{neq_comp_def_td} as charges are bound to particles. However, since inert charge do not effect particle dynamics, the particle full-counting statistics involves averaging solely over the particle initial measure \eqref{neq_def_td}
\begin{equation}
	G_p(\lambda_p|t) = \langle e^{\lambda_p J_p(t)} \rangle_{\mathcal{Q}_p}. \label{Gp_inert}
\end{equation}
Noting that the integrated particle current $J_p$ is invariant under arbitrary permutations of particle labels, the particle full-counting statistics of a single-file dynamics with elastic scattering is identical to that of a crossing dynamics by considering quasi-particles (unbroken straight lines in Fig.~\ref{fig:sf}) instead of particles as e.g.~in the hard-rod gas or generalized hydrodynamics \cite{Doyon_lectures}. In particular, even a free particle dynamics gives an interacting dynamics upon imposition of the single-file condition, which was used in \cite{Krajnik2022SF,Krajnik2022} to derive the exact finite-time particle and charge full-counting statistics for $\mathbb{Z}_2$ charges.

\section{Dressing}
\label{sec:dressing}
Charge current fluctuations can be obtained from fluctuations of the particle current by a combinatorial dressing procedure due to charge inertness \eqref{Uc_def} as we discuss below. The general form of the dressing simplifies for single-file dynamics \eqref{cond2}, resulting in a tractable integral representation.

The joint charge-particle distribution $\mathcal{P}_{c, p}$ of an arbitrary (i.e.~not necessarily inert or single-file) charged particle dynamics can be factorized into a product of the particle distribution and a time-independent conditional probability,
$	\mathcal{P}_{c,p}({J}_c,J_p|t) =  \mathcal{P}_{c|p}({J}_c|{J}_p) \mathcal{P}_p({J}_p|t). \label{joint_factor}$
The conditional distribution gives the probability of the integrated charge current $J_c$ conditioned on a value of the integrated particle current $J_p$. For a particle dynamics with inert charges the conditional probability is given by
\begin{widetext}
	\begin{equation}
		\mathcal{P}_{c|p}({J}_c|{J}_p) = \sum_{n_+, n_- = 0}^{\infty} \delta_{n_- - n_+, {J}_p} \int_{\mathcal{C}^{n_+ + n_-}} \dd \omega_{\mu_-}^{n_-} \dd \omega_{\mu_+}^{n_+}\ \delta(q_c^- - q_c^+ - {J}_c), \label{general_combinatorics}
	\end{equation}
\end{widetext}
where $n_\pm$ denote the number of particles crossing from right-to-left and vice-versa and $q_c^{\pm} \equiv \sum_{m = 1}^{n_\pm} c_{m}$ is the total charge carried by the respective crossing particles. Some comments on Eq.~\eqref{general_combinatorics} are in order. Since inert charges do not influence particle dynamics, the conditional probability is obtained by first summing over all combinations of particles crossing the origin (first sum and $\delta$-function in Eq.~\eqref{general_combinatorics}) and then averaging over possible charge configurations of these particles  (second $\delta$-function in Eq.~\eqref{general_combinatorics}). Given that charges are inert and not correlated in the initial measure \eqref{neq_comp_def_td}, this amounts to averaging over independent charge distributions (factorized charge measure in Eq.~\eqref{general_combinatorics}).

For a single-file dynamics \eqref{cond2}, either all particle cross from left to right or vice versa, so that at most one of $n_\pm \neq 0$ which simplifies the conditional probability \eqref{general_combinatorics}
\begin{equation}
	\mathcal{P}_{c|p}({J}_c|{J}_p) = \int_{\mathcal{C}^{|J_p|}}\dd \omega_{\mu_\nu}^{|J_p|} \delta(q_c^\nu + \nu {J}_c). \label{spin_combinatorics2}
\end{equation}
where $\nu = -{\rm sgn}({J}_p)$. The conditional probability is then (up to a sign) the distribution of the sum of $|J_p|$ charges sampled from $\omega_\mu$. 
By the convolution theorem, the conditional distribution reads
\begin{equation}
	\mathcal{P}_{c|p}({J}_c|{J}_p) = \int_\mathbb{R} \frac{\dd k}{2\pi} \hat \omega^{|J_p|}_{\mu_\nu}(k) e^{-\ii k \nu J_c}, \label{sum_dist}
\end{equation}
where we have introduced the Fourier transform of the charge measure
$\hat \omega_\mu(k) \equiv \int_{\mathcal{C}}\dd \omega_\mu\, e^{-\ii k c}.$
We denote the dressing of the particle distribution via factorization of the joint probability distribution and the conditional distribution \eqref{sum_dist} as the action of a dressing operator $\mathfrak{D}_{\mathcal{P}}$
\begin{equation}
	\mathcal{P}_{c, p}(J_c, J_p|t) = \mathfrak{D}_{\mathcal{P}}[\mathcal{P}_p(J_p|t)]. \label{p_dressing_def}
\end{equation}
The charge distribution $\mathcal{P}_c$ is recovered by marginalization $\mathcal{P}_c(J_c|t) = \int \dd J_p\, P_{c|p}(J_c|J_p)\mathcal{P}_p(J_p|t)$.

\paragraph*{{\bf Example}: Gaussian charges} An exactly solvable example of dressing is given by a Gaussian charge measure
\begin{equation}
	\omega_\mu^{\rm Gauss}(c) = \mathcal{N}_c(\mu, \sigma_\omega^2) \label{Gauss_measure},
\end{equation}
where $\mathcal{N}_x(\overline x, \sigma^2) \equiv e^{-(x-\overline x)^2/2\sigma^2}/\sqrt{2\pi \sigma^2}$, owing to additivity of variances of independent Gaussian variables.
 The  Fourier transform of Eq.~\eqref{Gauss_measure} reads
\begin{equation}
	\hat \omega_\mu^{\rm Gauss}(k) = 2\pi \mathcal{N}_\mu(0, \sigma_\omega^2) \mathcal{N}_k(-\ii\mu \sigma_\omega^{-2}, \sigma_\omega^{-2}). 
	\label{Gauss_measure_Fourier}
\end{equation}
Eq.~\eqref{sum_dist} reduces to a Gaussian integral which gives the exact conditional probability
\begin{equation}
	\mathcal{P}^{\rm Gauss}_{c|p}({J}_c|{J}_p) = \mathcal{N}_{J_c}(J_p \mu, |J_p|\sigma_\omega^2).
\end{equation}
Note that while Eq.~\eqref{sum_dist} gives the exact conditional probability for an arbitrary charge measure, the integral generally cannot be evaluated in closed form as for a Gaussian measure and must instead be evaluated asymptotically, see Section~\ref{sec:asymptotics}.

\subsection{Dressing the full-counting statistics}
A more compact and explicit relation between charge and particle fluctuations is obtained at the level of full-counting statistics. We define the action of the dressing operator on the particle full-counting statistics $\mathfrak{D}_{G}$ by analogy to the action of the dressing operator $\mathfrak{D}_\mathcal{P}$ \eqref{p_dressing_def}
\begin{equation}
	G_{c,p}(\lambda_c,\lambda_p|t) = \mathfrak{D}_{G}[G_p(\lambda_p|t)].
\end{equation}
Recalling the relation between the full-counting statistics and the probability distribution in terms of the Laplace transform \eqref{Laplace_P_G}, the dressing operators are related via conjugation by the Laplace transform
\begin{equation}
	\mathfrak{D}_{G} = \mathcal{L} \circ \mathfrak{D}_{\mathcal{P}} \circ \mathcal{L}^{-1}. \label{D_D_rel}
\end{equation}
The conjugation is computed in Appendix \ref{app:MGF_dressing}, resulting in a straightforward substitution rule for exponentials of the particle counting field
\begin{equation}
	\mathfrak{D}_G: e^{\pm n \lambda_p} \mapsto e^{\pm n \lambda_p}  \hat \omega_{\mu_\mp}^n(\pm \ii \lambda_c), \quad n \in \mathbb{N}. \label{Dg_rule}
\end{equation}
We note that $\hat \omega_{\mu_\mp}(\pm \ii \lambda_c) = \int_{\mathcal{C}} \dd \omega_{\mu_\mp} e^{\pm c \lambda_c}$ are the charge full-counting statistics of a single particle going left-to-right or right-to-left respectively. Since charges are uncorrelated in the initial measure \eqref{neq_comp_def}, the corresponding $n$-particle result factorizes in terms of the single-particle one. The substitution rule \eqref{Dg_rule} therefore dresses the particle full-counting statistics by adjoining the left/right-going charge full-counting statistics with fixed particle number to the corresponding particle crossings.
\paragraph*{{\bf Example}: $\mathbb{Z}_2$ charges} 
{The $\mathbb{Z}_2$ charge measure reads
	\begin{equation}
		\omega^{\mathbb{Z}_2}_\mu(c) = \frac{1- \mu}{2} \delta(c+1) + \frac{1+ \mu}{2} \delta(c-1), \label{Z2_measure}
	\end{equation}
	with the Fourier transform $\hat \omega^{\mathbb{Z}_2}_\mu(k) = \cos k -\ii \mu \sin k$. The dressing substitution 
	\begin{equation}
		\mathfrak{D}_G^{\mathbb{Z}_2}: e^{\pm n \lambda_p} \mapsto e^{\pm n \lambda_p}  [\cosh \lambda_c \pm \mu_\mp \sinh \lambda_c]^n
	\end{equation}
	recovers the corresponding result for the $\mathbb{Z}_2$ dressing operator in \cite{Krajnik2022SF}.
}

\section{Dressing asymptotics}
\label{sec:asymptotics}
The dressing operator \eqref{Dg_rule} gives the exact finite-time dressing of particle full-counting statistics. We now consider how the dressing manifests in time-asymptotic charge fluctuations on typical and large scales which we specify below. We show that typical fluctuations depend only on the first two moments of the charge measure. Assuming asymptotically Gaussian typical particle fluctuations, we demonstrate that typical charge fluctuations in equilibrium ensembles with vanishing mean charge are \emph{anomalous} (non-normal).\\

Gaussianity of typical fluctuations is commonly established as a consequence of a large deviation principle \cite{Touchette_LDT}. To proceed, we make the following assumptions about the particle full-counting statistics:
\begin{enumerate}[label=(\roman*)]
\item The particle full-counting statistics satisfy a large deviation principle with a strictly convex twice differentiable particle scaled cumulant generating function $F_p \in C^2(\mathbb{R})$
\begin{equation}
	F_p(\lambda_p) \equiv \lim_{t \to \infty}t^{-\alpha} \log G_p(\lambda_p|t), \label{Fp_def}
\end{equation}
where the speed exponent $\alpha \in \mathbb{R}_+$ is chosen such that $F_p$ is finite and non-zero.

\item Asymptotic values of scaled cumulants $s_n^{(p)} \equiv \lim_{t \to \infty} t^{-\alpha} \partial^n_{\lambda_p} \log G_p(\lambda_p|t)|_{\lambda_p = 0}$ are finite, $|s_n^{(p)}| < \infty$, for all $n \in \mathbb{N}$ and the asymptotic value of the second cumulant is nonzero, $\sigma_p^2 \equiv s_2^{(p)} \neq 0$.
\end{enumerate}

The asymptotic exponential decay of \emph{large} (of order $\mathcal{O}(t^{\alpha})$) fluctuations of the integrated particle current $J_p$ is governed by a large deviation principle \cite{Touchette_LDT} \begin{equation}
	\mathcal{P}_p(J_p = j_p t^{\alpha} |t) \asymp e^{-t^{\alpha}I_p(j_p)}, \label{particle_LDP}
\end{equation}
where $I_p(j_p) \equiv -\lim_{t \to \infty} t^{-\alpha} \log \mathcal{P}_p(J_p = j_p t^{\alpha} |t) $ is the particle rate function and $\asymp$ indicates asymptotic logarithmic equivalence. The twice
 differentiable $F_p(\lambda_p)$ in assumption (i) is related to the rate function $I_p$ is related by the Legendre-Fenchel transform
\begin{equation}
	I_p(j_p) = \sup_{\lambda \in \mathbb{R}} \{\lambda_p j_p -F_p(\lambda_p)\}, \label{Ip_min}
\end{equation}
which shows that $I_p$ is a twice differentiable strictly convex function on its domain $\mathcal{J}_p \equiv (j_p^{\rm min}, j_p^{\rm max})$. 

While large fluctuations become exponentially unlikely at large times, \emph{typical} (of order $\mathcal{O}(t^{\alpha/2})$) fluctuations have a finite asymptotic distribution. Noting that the average integrated particle current is given asymptotically by the first cumulant $\langle J_p(t) \rangle \simeq  s_1^{(p)} t^{\alpha}$ we denote $J_p[j_p] = s_1^{(p)} t^{\alpha} + {j}_p t^{\alpha/2}$. The asymptotic typical distribution is then given by
\begin{equation}
	\mathcal{P}^{\rm typ}_p(j_p) \equiv \lim_{t \to \infty} t^{\alpha/2} \mathcal{P}_p(J_p=J_p[j_p]|t). \label{particle_typ_def}
\end{equation}
Taken together, assumption (i) and (ii) ensure that the typical distribution of $j_p$ is asymptotically normal
\begin{equation}
	\mathcal{P}^{\rm typ}_p(j_p) = \mathcal{N}_{j_p}(0, \sigma_p^2). \label{particle_typ}
\end{equation}
We note that finiteness of scaled cumulants in (ii) and the resulting asymptotic normality generally does not follow directly from the existence of the limit in (i). Instead, a sufficient condition is that the logarithm of the particle full-counting statistics $\log G_p(\lambda_p|t)$ is \emph{complex}-analytic at the origin $\lambda_p=0$ for all $t$, see \cite{Bryc1993}. The assumption (ii) should be understood as a regularity condition on the particle dynamics that ensures central limit behavior on the typical scale set by the variance.

\subsection{Typical charge fluctuations}

Consider a non-equilibrium initial ensemble \eqref{neq_comp_def_td} with generic particle densities $\rho_+ \neq \rho_-$ and charge means $\mu_\pm \neq 0$. The difference of particle densities drives average particle and charge currents $s_1^{(i)} \neq 0$. To access typical fluctuations around the averages, we set
$J_i[j_i] = s_1^{(i)} t^{\alpha} + {j}_i t^{\alpha/2}$
and consider the joint typical distribution
$\mathcal{P}^{\rm typ}_{c, p}({j}_c, {j}_p)  \equiv \lim_{t \to \infty} t^{\alpha} \mathcal{P}_{c, p}(J_c[j_c], J_p[j_p]|t).$
The factorized form of the joint typical distribution at the typical scale then reads $\mathcal{P}^{\rm typ}_{c, p}({j}_c, {j}_p) = \mathcal{P}^{\rm typ}_{c|p}({j}_c|{j}_p) \mathcal{P}^{\rm typ}_{p}({j}_p)$,
where the typical conditional probability is
$\mathcal{P}^{\rm typ}_{c|p}({j}_c|{j}_p) \equiv \lim_{t \to \infty} t^{\alpha/2} \mathcal{P}_{c|p}(J_c[j_c]| J_p[j_p]).$
As shown in Appendix \ref{app:typ_cond1}, the integral representation \eqref{sum_dist} allows us to relate the average values of charge and particle currents, giving the expected result 
\begin{equation}
	s_1^{(c)} = s_1^{(p)} \mu_{\overline \nu}, \label{avg_c_p}
\end{equation}
where $\overline \nu = - {\rm sgn}(s_1^{(p)})$ with Gaussian fluctuations around the mean value,
$\mathcal{P}^{\rm typ}_{c|p}({j}_c|{j}_p) = \mathcal{N}_{j_c}(j_p \mu_{\overline \nu}, |s_1^{(p)}|\sigma_\omega^2).$
The typical charge distribution is obtained by marginalization, yielding a Gaussian distribution
\begin{equation}
	\mathcal{P}^{\rm typ}_{c}({j}_c)  = 
	\mathcal{N}_{j_c}\left(0, \sigma_c^2\right), \label{typ_charge_gauss}
\end{equation} 
with charge variance $\sigma_c^2 = \mu^2_{\overline \nu}\sigma_p^2 + |s_1^{(p)}|\sigma^2_\omega.$
The charge variance decomposes into a contribution stemming from typical particle fluctuations with variance $\sigma_p^2$ and average charge $\mu_{\overline \nu}$ and a contribution from the charge measure variance $\sigma_\omega^2$ carried by the average particle current $|s_1^{(p)}|$.

In equilibrium ensembles with vanishing mean charge, $\rho_\pm = \rho$, $\mu_\pm = 0$, the average particle and charge currents vanish $s_1^{(p)}=s_1^{(c)}=0$, see Eq.~$\eqref{avg_c_p}$. Moreover, the vanishing charge variance $\sigma_c=0$ indicates that the distribution \eqref{typ_charge_gauss} approaches a $\delta$-function centered at $j_c=0$. Typical charge fluctuation in such ensembles are not of order $\mathcal{O}(t^{\alpha/2})$ but are instead $\mathcal{O}(t^{\alpha/4})$. We accordingly define the typical conditional probability in equilibrium ensembles with vanishing mean charge$
	\mathcal{P}^{\rm typ,\,eq}_{c|p}({j}_c|{j}_p) \equiv \lim_{t \to \infty} t^{\alpha/4} \mathcal{P}_{c|p}(J_c=j_ct^{\alpha/4}| J_p(j_p))$.
As shown in Appendix \ref{app:typ_cond2}, Eq.~\eqref{sum_dist} now gives $\mathcal{P}^{\rm typ,\,eq}_{c|p}({j}_c|{j}_p) = \mathcal{N}_{j_c}(0, |j_p|\sigma_\omega^2)$.

 \begin{figure}[t!]
 	\includegraphics[width=\linewidth]{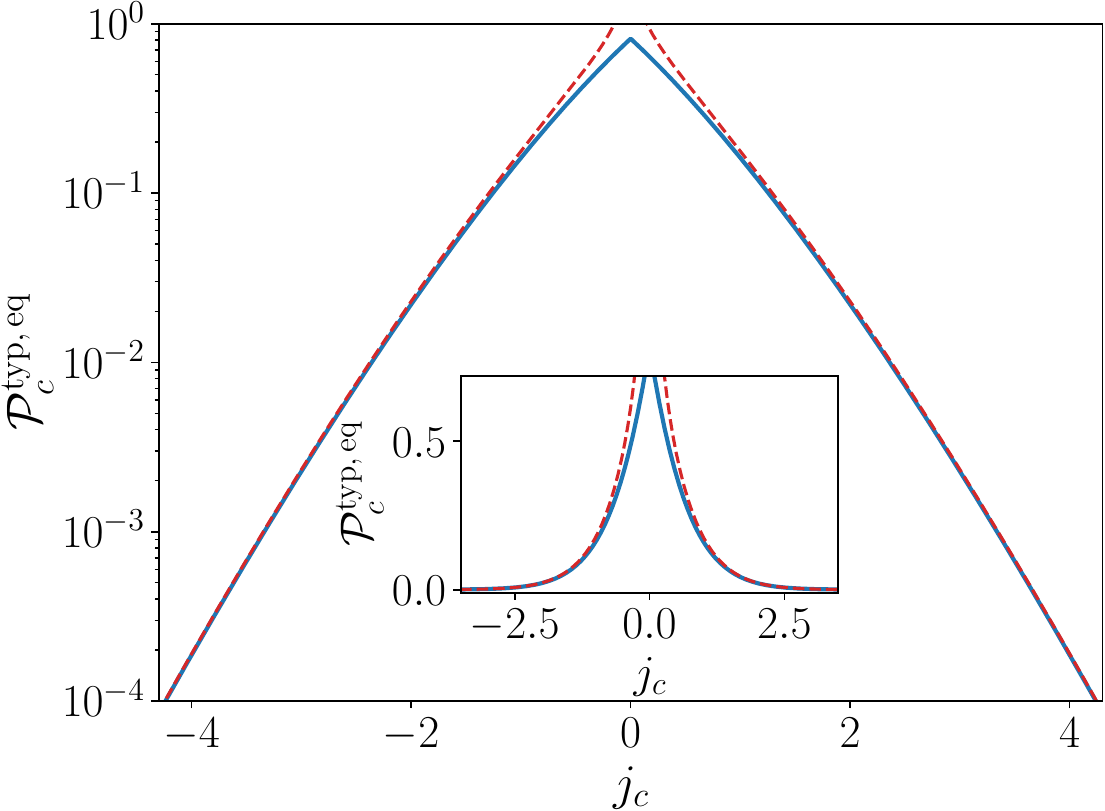}
 	\caption{Typical distribution of transported charge $\mathcal{P}_{c}^{\rm typ,\, eq}$ \eqref{typ_charge_wright} with $\sigma=1$ (full blue curves) in an equilibrium ensemble with vanishing mean charge in logarithmic (main figure) and linear (inset) scale. The distribution is strongly non-Gaussian, decaying asymptotically as $\mathcal{P}_{c}^{\rm typ,\, eq}(\sigma j_c/\sqrt{2}) \simeq  2|j_c|^{-1/3}e^{-\tfrac{3}{4}|j_c|^{4/3}}/\sqrt{3\pi}$ (dashed red curves) and featuring a cusp at $j_c=0$.}
 	\label{fig:Wright}
 \end{figure} 

Remarkably, the resulting typical charge distribution $\mathcal{P}^{\rm typ, eq}_c(j_c) = \int_{\mathbb{R}} \dd j_p\, \mathcal{P}^{\rm typ,\,eq}_{c|p}({j}_c|{j}_p) \mathcal{P}^{\rm typ}_{p}({j}_p)$ is non-Gaussian but is instead given by an M-Wright distribution, see Ref.~\cite{Mainardi_2020}, Appendix \ref{app:MWright} and Figure \ref{fig:Wright}
\begin{align}
	\mathcal{P}^{\rm typ, eq}_{c}({j}_c) =  \sigma^{-1}M_{1/4}(2|j_c|/\sigma), \label{typ_charge_wright}
\end{align}
where $M_{1/4}(|x|) =  \int_{\mathbb{R}} \frac{\dd y}{2\pi}|y|^{-1/2} e^{-x^2/4|y| - y^2/4}$ and $\sigma^2\equiv \sqrt{2} \sigma_p\sigma_\omega^2$.
Compared to a Gaussian distribution, the anomalous distribution \eqref{typ_charge_wright} has heavier tails with the leading asymptotic term of the M-Wright function given by Laplace's method as $M_{1/4}(\sqrt{2}(2x)^3) \simeq \tfrac{e^{-12x^{4}}}{\sqrt{3\pi x^2}} $ for $|x| \to \infty$.
M-Wright functions of general order occur as fundamental solutions in stochastic processes described by fractional diffusion equations \cite{Mainardi1996,Mainardi_2020,Barkai_2002,PhysRevE.61.132}. 
 In the realm of fluctuations in deterministic systems, the distribution \eqref{typ_charge_wright} has already been found to govern charge fluctuations in inert $\mathbb{Z}_2$ charged single-file systems \cite{Krajnik2022,Krajnik2022SF}. Unlike in fractional diffusion equations, the order of the M-Wright function in single-file systems is fixed to $1/4$ independently of the time-scale of underlying dynamics. 
 Our result shows that anomalous equilibrium typical fluctuations are independent of the detailed structure of the charge measure, requiring only a finite variance and the single-file property \eqref{cond2}.

\subsection{Large charge fluctuations}
Typical charge fluctuations depend only on the first two moments of the charge measure, whereas large fluctuations are sensitive to the full structure of the measure. In the following it will be convenient that the Fourier transform of the measure $\hat \omega_\mu(k)$ is an entire function. By the Paley-Wiener theorem \cite{Rudin_book}, a sufficient assumption is that the charge measure decays faster than exponentially on the real line, $\log(\omega_\mu(c))/|c| \rightarrow - \infty$ as $|c|\to \infty$.

The joint charge-particle rate function $I_{c, p}(j_c,j_p) = -\lim_{t \to \infty} t^{-\alpha} \log \mathcal{P}_{c, p}(J_c= j_c t^\alpha, J_p = j_p t^\alpha)$ is obtained by taking the logarithm of the factorized probability
\begin{equation}
	I_{c, p}(j_c, j_p) = I_{c|p}(j_c|j_p) + I_p(j_p), \label{Icp_def}
\end{equation}
where $I_{c|p}$ is the conditional rate function given by
\begin{equation}
	I_{c|p}(j_c|j_p) \equiv -\lim_{t \to \infty} t^{-\alpha} \log  \int_{\mathbb{R}} \frac{\dd k}{2\pi} \hat \omega^{|j_p|t^\alpha}_{\mu_\nu}(k)e^{-\ii \nu k j_c t^{\alpha}}. \label{cond_icp_int}
\end{equation}
The rate function $I_c(j_c) \equiv -\lim_{t \to \infty} t^{-\alpha} \log P_c(J_c = j_c t^\alpha |t)$ is recovered by marginalizing the joint rate function \eqref{Icp_def} over all particle fluctuations which reduces to a minimization problem
\begin{equation}
	I_c(j_c) = \min_{j_p \in \mathcal{J}_p^- \cup \mathcal{J}_p^+} I_{c, p}(j_c, j_p), \label{Ic_minimization}
\end{equation}
over the domains $\mathcal{J}_p{^\pm}$ of the particle current, see Appendix \ref{app:large_cond} for details.

\paragraph*{{\bf Example}: Large fluctuations of Gaussian charges}
Gaussian charges \eqref{Gauss_measure} again provide a useful exactly solvable example with the following representation for the conditional rate function
\begin{equation}
	I_{c|p}^{\rm Gauss}(j_c|j_p) = -\lim_{t \to \infty} t^{-\alpha} \log  \int_{\mathbb{R}} \frac{\dd k}{2\pi} e^{t^\alpha f(k)}, \label{Icp_Gauss1}
\end{equation}
where
$f(k) = -\frac{1}{2} |j_p| \sigma_\omega^2 k^2 - \ii k\nu(j_c - j_p \mu). $
The rate function follows from a Gaussian integral
\begin{equation}
	I_{c|p}^{\rm Gauss}(j_c|j_p) = \frac{(j_c-j_p \mu_\nu)^2}{2\sigma_\omega^2 |j_p|}. \label{Icp_Gauss3}
\end{equation}
The simplicity of the result \eqref{Icp_Gauss3} allows it to serve as a tractable example in the following more involved analysis of the general case, which nevertheless contains all the main intricacies of the analysis.

\subsubsection{First minimization problem}
We analyze the minimization problem \eqref{Ic_minimization} by splitting the conditional rate function according to the sign of the particle current
\begin{equation}
	I_{c|p}(j_c|j_p) = 
	\begin{cases}
		I_{c|p}^{+}(j_c|j_p) & {\rm for}\ j_p \in \mathcal{J}_p^{-},\\
		I_{c|p}^{-}(j_c|j_p) & {\rm for}\ j_p \in \mathcal{J}_p^{+},
	\end{cases}
\end{equation}
and decompose \eqref{Ic_minimization} into two sequential minimizations.
We first minimize $I^\pm_{c, p}(j_c, j_p) \equiv I_{c|p}^\pm(j_c|j_p) + I_p(j_p)$  over $\mathcal{J}_p^\mp$ to obtain the split charge rate functions $I_c^\pm$
\begin{equation}
	I_c^{\pm}(j_c) = \min_{j_p \in \mathcal{J}_p^{\mp}}I^\pm_{c, p}(j_c, j_p). \label{first_optimization}
\end{equation}
The split conditional rate functions $I_{c|p}^{\pm}$ are computed in Appendix \ref{app:LD_asymptotics}, resulting in
\begin{equation}
	\pm I_{c|p}^{\pm}(j_c|j_p)  = j_p \log \hat \omega_{\mu_\pm}(\ii \kappa_\pm) - j_c \kappa_\pm \quad j_p \in \mathcal{J}_p^{\mp},
	\label{split_cond_rate_result}
\end{equation}
 with  $\kappa_\pm \in \mathbb{R}$ the unique solutions of saddle point equations on the imaginary axis
\begin{equation}
	j_p\, \partial_k \hat\omega_{\mu_\pm}(k)|_{k=\ii \kappa_\pm} + \ii j_c\, \hat \omega_{\mu_\pm}\left(\ii \kappa_\pm\right) = 0.
	\label{saddle_point_eq}
\end{equation}
\begin{figure}[t!]
	\includegraphics[width=\linewidth]{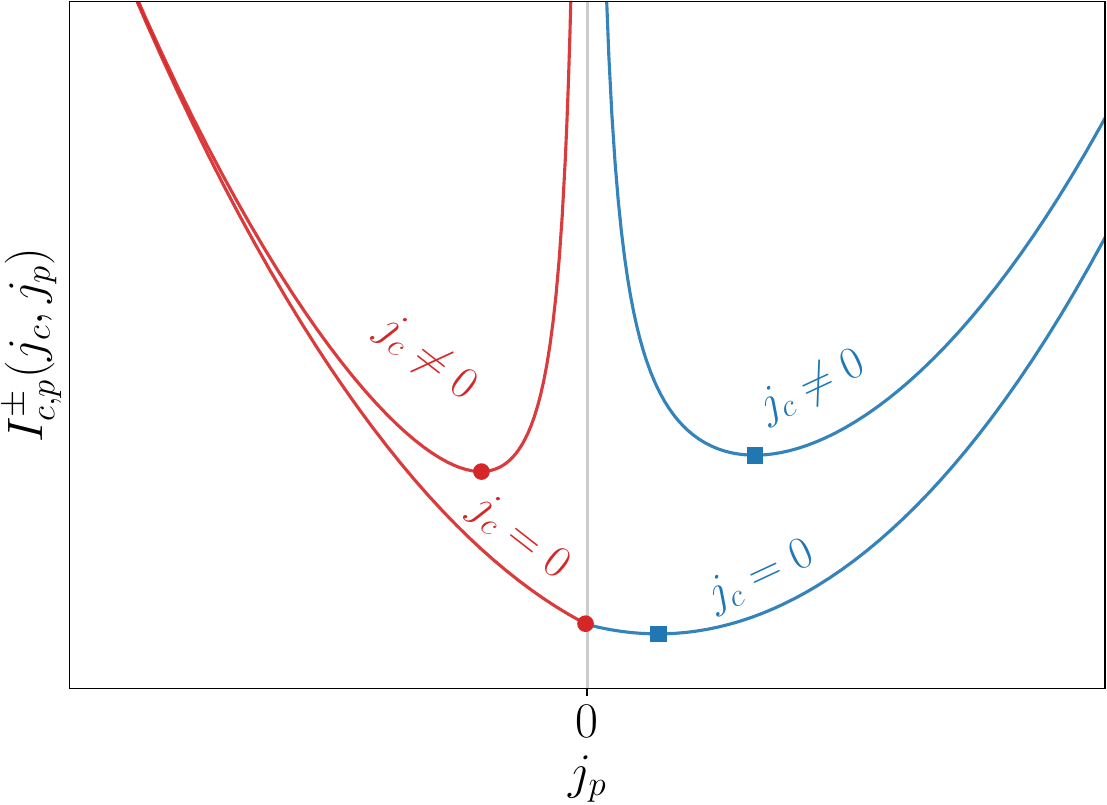}
	\caption{Split rate functions $I_{c,p}^\pm$ (red/left and blue/right lines respectively) for a Gaussian charge measure \eqref{Gauss_measure} and their respective minima at $j_p = j_{\rm crit}^\pm(j_c)$ \eqref{jcrit_def} (red circles/blue squares). For $j_c\neq0$ the functions $I_{c,p}^\pm$ diverge near the origin by \eqref{statement3}. For $j_c=0$, $I_{c,p}^\pm$ are continuous and convex as a function of $j_p$ at the origin due to \eqref{statement1} and \eqref{statement2} respectively. Note the minimum of $I_{c,p}^+$ at $j_p=0$ for $j_c=0$.}
	\label{fig:2}
\end{figure}

\begin{figure*}[t!]
	\includegraphics[width=\textwidth]{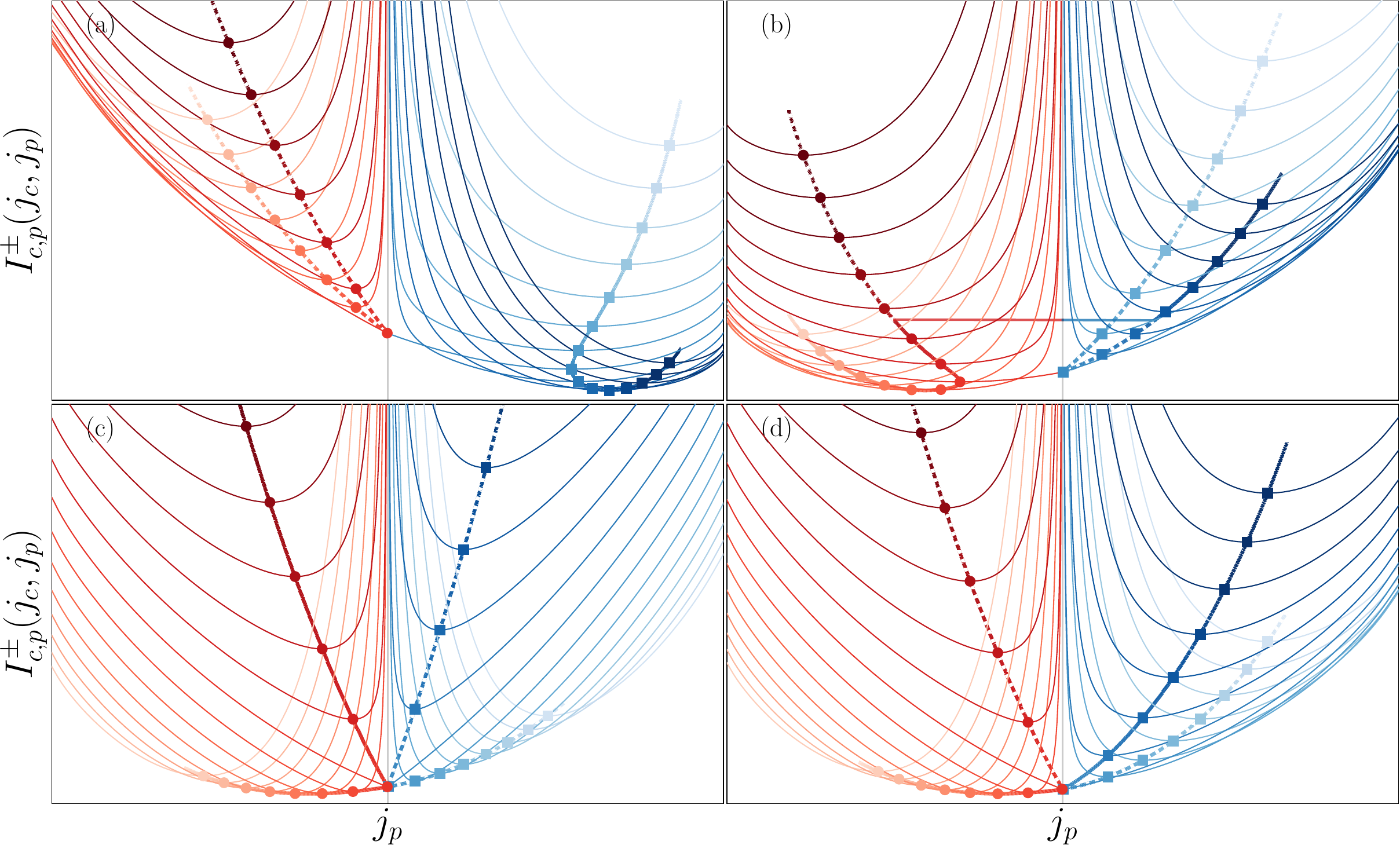}
	\caption{Two-step minimization of split rate functions $I_{c,p}^\pm$ (thin red/blue lines on left/right of each panel respectively) for a Gaussian charge measure \eqref{Gauss_measure}. The first minimization \eqref{first_optimization} gives minima (red/blue circles) at $j_p = j_{\rm crit}^\pm(j_c)$ \eqref{jcrit_def} (red circles/blue squares) that follow dashed thick red/blue (left/right) lines  as functions of $j_c$. The flow of global minima - solutions of \eqref{second_optimization} - is indicated by solid thick red/blue (left/right) lines and displays qualitatively different behavior upon varying parameters of the initial ensemble \eqref{neq_comp_def_td}.}
	\label{fig:3}
\end{figure*}

While Eq.~\eqref{split_cond_rate_result} cannot be simplified further without additionally specifying the measure, it suffices to demonstrate the following general properties of split rate functions in their respective domains, see Appendix \ref{app:generic_properties} and Figures~\ref{fig:2} and \ref{fig:3}:
\begin{enumerate}[label=(\arabic*)]
	\item $I_{c,p}^{\pm}(j_c=0,j_p)$ approach $I_p(j_p)$ as $j_p \to 0^\mp$
	\begin{equation}
		\lim_{j_p \to 0^\mp} I_{c|p}^{\pm}(j_c = 0, j_p) = I_p(j_p). \label{statement1}
	\end{equation} 
	\item Derivatives of $I_{c,p}^{\pm}(j_c=0,j_p)$ with respect to $j_p$ are finite and bounded by
	\begin{equation}
		\mp \infty < \pm \partial_{j_p} I_{c,p}^{\pm}(j_c = 0, j_p \in \mathcal{J}_p^\mp ) \leq   \pm \partial_{j_p}I_p(j_p). \label{statement2}
	\end{equation}
	\item Derivatives of $I_{c,p}^{\pm}(j_c\neq 0,j_p)$ with respect to $j_p$ diverge on inner boundaries of $\mathcal{J}_p^\pm$
	\begin{equation}
		\pm \partial_{j_p} I_{c,p}^{\pm}(j_c \neq 0| j_p \to j_p^\mp) \to  \infty. \label{statement3}
	\end{equation}
	\item The Hessian determinant of $I_{c,p}^\pm(j_c,j_p)$ is positive
	\begin{equation}
		\det {\rm H}_{I_{c,p}^\pm} (j_c,j_p) > 0. \label{statement4}
	\end{equation}
\end{enumerate}

Having established these general properties of the split rate functions, we are in a position to describe the structure of the split charge rate functions $I_c^\pm$ resulting from the minimization \eqref{first_optimization}. Note that the properties (\ref{statement1}-\ref{statement4}) are similar to those used in the analysis of the minimization problem \eqref{first_optimization} for a $\mathbb{Z}_2$ charge measure \eqref{Z2_measure} in Ref.~\cite{Krajnik2022SF}.\\

We start by analyzing the optimization problem \eqref{first_optimization} for $j_c \neq 0$. 
Positivity of the Hessian determinant \eqref{statement4} means that $I^\pm_{c, p}$ are strictly convex functions that
diverge near the inner boundaries of $\mathcal{J}_p^\mp$ by \eqref{statement3}. Since
their derivatives change sign on $\mathcal{J}_p^{\mp}$, $I_{c,p}^\pm$ each has a single minimum at $j_p=j^\pm_{\rm crit}(j_c)$ in the interior of the domains $\mathcal{J}_p^{\mp}$, see Fig.~\ref{fig:2}, where the derivatives of $I^\pm_{c, p}$ vanish
\begin{equation}
	\partial_{j_p} I^\pm_{c,p}(j_c \neq 0, j_p)|_{j_p = j^\pm_{\rm crit}}.  \label{jcrit_def}
\end{equation}
The split charge rate functions are then given by
\begin{equation}
	I_c^\pm(j_c) = I^\pm_{c,p} (j_c, j_p)|_{j_p=j^\pm_{\rm crit}(j_c)} \label{Icpm_result}
\end{equation}
and are strictly convex since minimization over the domains $\mathcal{J}_p^\mp$ preserves strict convexity of $I_{c,p}^\pm$.
By the implicit function theorem and strict convexity, the minima $j^\pm_{\rm crit}$ are differentiable functions of $j_c$. Differentiability of $I^\pm_{c}$ for $j_c \neq 0$ finally follows by noting that $I^\pm_{c,p}$ are differentiable in both arguments.

It remains to consider the optimization problem \eqref{first_optimization} for $j_c=0$. 
The derivatives of $I^\pm_{c,p}$ at $j_p=0$ are finite by \eqref{statement2} and the minima of $I^\pm_{c,p}$ are no longer guaranteed to occur in the interior of $\mathcal{J}_p^\mp$, and can instead approach the origin, see Fig.~\ref{fig:2}, resulting in two possible cases 
\begin{equation}
	\lim_{j_c \to 0}I^\pm_{c}(j_c) =
	\begin{cases}
		I^\pm_{c, p}(j_c, j^\pm_{\rm crit}(j_c) \neq 0),\\
		I^\pm_{c, p}(j_c, j_p = 0) = I_p(0). \label{jc0_cases}
	\end{cases}
\end{equation}
In the first case $I_c^\pm$ are differentiable on their entire domains.
In the second case $I^\pm_{c}$ are non-differentiable at $j_c=0$ due to a violation of the condition for an interior minimum \eqref{jcrit_def}, but remain strictly convex by virtue of strict convexity of $I^\pm_{c,p}$ \eqref{statement4}.


The domains $\mathcal{J}_p^\pm$ join at $j_c=0$, $\lim_{j_c \to 0} \mathcal{J}_p^-(j_c) \cup \mathcal{J}_p^+(j_c) = \mathcal{J}_p$  and $I_{c,p}(j_c=0, j_p)$ is continuous on $\mathcal{J}_p$ by \eqref{statement1}. However, the right/left derivatives of $I^\pm_{c,p}(j_c=0,j_p)$ at $j_p=0$ in general do not match, rendering $I_{c,p}$ non-differentiable at the origin. Instead continuity at the origin and \eqref{statement3} make $I_{c,p}$ strictly convex on $\mathcal{J}_p$, while its derivative changes sign on $\mathcal{J}_p$. As such, it has only a single minimum on $\mathcal{J}_p$ so that at least one of the minima of $I^\pm_{c,p}(j_c=0, j_p)$ are necessarily located at $j_p=0$, see Fig.~\ref{fig:2},  and the corresponding solution $I_c^\pm$ is non-differentiable at $j_c=0$.

To summarize, the split charge rate functions $I^\pm_c$, obtained as solutions of \eqref{first_optimization} are strictly convex differentiable functions, except at the origin, where at least one of the functions is non-differentiable.

\subsubsection{Second optimization problem}
Having obtained the split charge rate function $I^\pm_{c}$, we recover the full charge rate function $I_c$ by simply picking the smaller of the two at each value of the charge current
\begin{equation}
	I_c(j_c) = \min \{I_c^{-}(j_c), I_c^{+}(j_c)\}. \label{second_optimization}
\end{equation}
The simplest case is that one of the split charge rate functions $I^\pm_c$ dominates the other on the entire charge current domain, trivially resulting in a differentiable charge rate function $I_c(j) = I_c^\pm(j_c)$, see Fig.~\ref{fig:3}a. However, nothing prevents one of the rate functions overtaking the other at a generic charge current $j_{c}^{\rm cusp}$, see Figs.~\ref{fig:3}b and \ref{fig1}a
\begin{equation}
	I^+_c(j_{c}^{\rm cusp}) = I^-_c(j_c^{\rm cusp}) \label{cusp},
\end{equation}
resulting in a non-differentiable point of the charge rate function $I_c$. Since $I_c^\pm$ are always differentiable away from the origin, the difference of right and left derivatives $\Delta^\pm (j_c) \equiv \partial_{\delta}\left[I_{c}^+(j_c \pm \delta) - I_c^-(j_c \pm \delta)\right]|_{\delta=0}$ changes sign at the cusp $\Delta^+(j_c^{\rm cusp}) \Delta^-(j_c^{\rm cusp}) = -\left(\partial_{j_c} [I^+_c(j_c) - I^-_c(j_c)]|_{j_c = j_c^{\rm cusp}} \right)^2 < 0$,
indicating that $I_c$ has a non-convex cusp at $j_c^{\rm cusp}$.

It remains to consider the possibility of crossing at the origin, which can occur when both $I_c^+$ and $I_c^-$ have a non-differentiable point at $j_c=0$ with $I^{\pm}_c(0) = I_p(0)$ by Eq.~\eqref{jc0_cases}. The right and left derivatives at the origin in general do not match
\begin{equation}
	I_c(j_c) =
	\begin{cases}
		I^\pm(j_c) \textrm{ for } j_c > 0 \enspace \textrm{if}\ \pm \Delta^+(0) < 0,\\
		I^\pm(j_c) \textrm{ for } j_c < 0 \enspace \textrm{if}\ \pm \Delta^-(0) > 0.
	\end{cases} \label{origin_nondif}
\end{equation}
We accordingly distinguish two cases.
In the first case, the same function $I_c^\pm$ is smaller on both sides of the origin, see Fig.~\ref{fig:3}c. The resulting $I_c$ is trivially non-differentiable at the origin and strictly convex in a neighborhood of the origin. In the second case, $I_c$ switches from $I^\pm_c$ to $I_c^\mp$ as it crosses the origin, see Figs.~\ref{fig:3}d and \ref{fig2}a. The resulting $I_c$ is again non-differentiable at the origin but is no longer necessarily convex and can exhibit a non-convex cusp.

To summarize, the charge rate functions $I_c$, obtained as solutions of the optimization problem \eqref{second_optimization}, can exhibit two basic types of non-differentiable behavior, exemplified in Fig.~\ref{fig:3}. The first type correspond to a non-convex cusp in the rate function, at which the two split rate functions exchange dominance. The second type is manifested as non-differentiability of the rate function at the origin. In the first case one of the split rate functions dominates on both sides of the origin the charge rate function is strictly convex. When split rate functions exchange dominance at the origin the charge rate function can exhibit a non-convex cusp at the origin.  We emphasize that the charge rate function can combine both a non-differentiable cusp  for $j_c^{\rm cusp}\neq 0$ with non-differentiability at the origin.


\subsection{Dynamical phase transitions}
The non-differentiable behavior of the charge rate function $I_c$ is more conveniently discussed by considering the charge scaled cumulant generating function
\begin{equation}
	F_c(\lambda_c) \equiv \lim_{t \to \infty}t^{-\alpha} \log G_c(\lambda_c|t), \label{Fc_def}
\end{equation}
commonly referred to as the dynamical free energy \cite{Oono_1989}, see also the discussion in \cite{Krajnik2022SF}.
\begin{figure}[h!]
	\includegraphics[width=\linewidth]{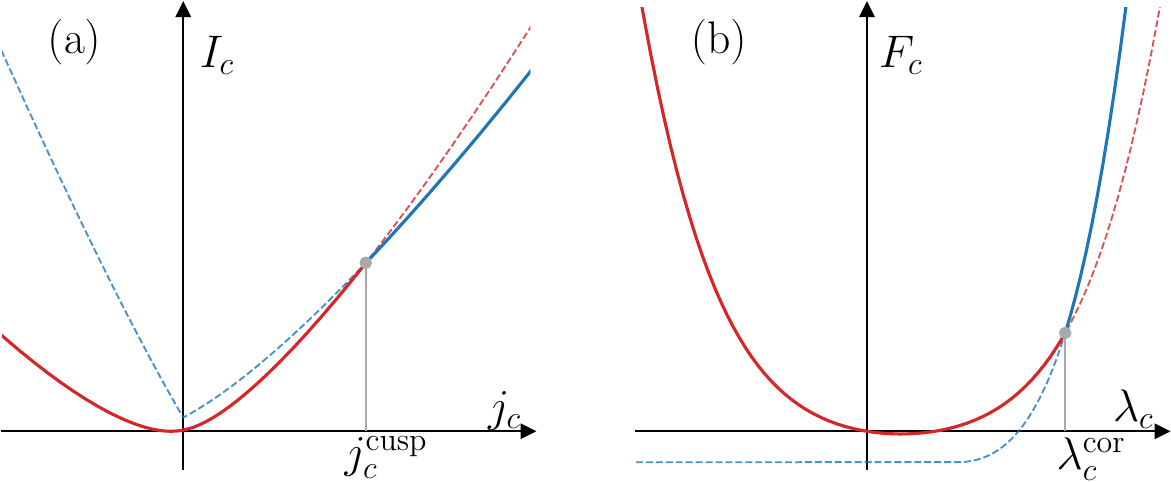}
	\caption{A first order dynamical phase transition. (a) Split charge rate functions $I_c^\pm$ (solid and dashed red/blue lines) entering the second minimization problem \eqref{second_optimization}. The charge rate function $I_c$ (solid line)  has non-convex cusp at $j_c^{\rm cusp}$ (gray circle). (b) Split charge scaled cumulant generating functions $F_c^\pm(\lambda_c)$ \eqref{Fpm} (solid and dashed red/blue lines) entering the second optimization problem \eqref{opt2_2}. The non-convex cusp of $I_c$ is mapped to a non-differentiable corner (gray circle) of the charge scaled cumulant generating function $F_c$ (solid line) at $\lambda_c^{\rm cor}$.}
	\label{fig1}
\end{figure}
Scaled cumulant generating functions are convex and recoverable from a rate function by the Legendre-Fenchel transform irrespective of its differentiability \cite{Touchette_LDT}
\begin{equation}
	F_c(\lambda_c) = \sup_{j_c} \{\lambda_c j_c - I_c(j_c)\}. \label{I_to_F}
\end{equation}
The two-step structure of the charge rate function minimization is inherited by the optimization problem \eqref{I_to_F}, leading to an initial optimization for the split charge scaled cumulant generating functions
\begin{equation}
	F^\pm_c(\lambda_c) = \sup_{j_c} \{\lambda_c j_c - I_c^\pm(j_c)\}, \label{Fpm}
\end{equation}
followed by picking the larger of the two results
\begin{equation}
	F_c(\lambda_c) = \max\{F_c^-(\lambda_c), F_c^+(\lambda_c) \}. \label{opt2_2}
\end{equation}
A non-convex cusp \eqref{cusp} at $j_c^{\rm cusp}$ (including at $j_c^{\rm cusp}=0$) of the charge rate function $I_c$ is accordingly mapped to a convex non-differentiable corner of $F_c$ at $\lambda_c^{\rm cor}$, see Figure~\ref{fig1}. From the miss-match of first derivatives of $F_c$ at the corner we conclude that a non-convex cusp of $I_c$ corresponds to a first-order dynamical phase transition.

\begin{figure}[h!]
	\includegraphics[width=\linewidth]{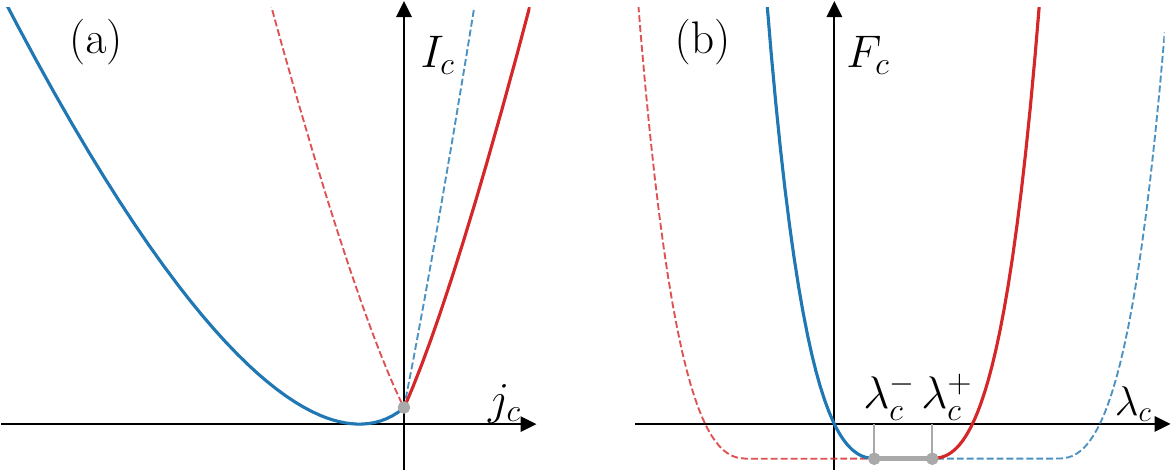}
	\caption{A second order dynamical phase transition. (a) A pair of split charge rate functions $I_c^\pm$ (solid and dashed red/blue lines)  exchange dominance at the origin $j_c=0$, where the strictly convex charge rate function $I_c$ (solid line)  has non-differentiable point (gray circle). (b) Split charge scaled cumulant generating functions $F_c^\pm(\lambda_c)$ \eqref{Fpm} (solid and dashed red/blue lines) exhibit a flat plateau. The non-convex point of $I_c$ at the origin is mapped to a flat interval \eqref{Fc_flat} (horizontal gray line) of the charge scaled cumulant generating function $F_c$ (solid line) with a discontinuous second derivative at the boundaries $\lambda_c^\pm$ (gray circles).}
	\label{fig2}
\end{figure}

On the other hand, a non-differentiable point at the origin of a strictly convex charge rate function $I_c$ \eqref{origin_nondif} is mapped to a flat interval of $F_c$, see Figure~\ref{fig2}, whose boundaries are given by the right and left derivatives of the rate function at the origin
$F_c(\lambda_c) = I_p(0) \quad {\rm for}\enspace  \lambda_c \in (\lambda_c^-,\ \lambda_c^+), \label{Fc_flat}
$
where $\lambda_c^\pm = \partial_{j_c}I_c(0^\pm)$.
By taking a derivative of Eq.~\eqref{I_to_F}, we find that outer derivatives of $F_c$ on the boundaries of the interval vanish, $\partial_{\lambda_c}F_c(\lambda_c)|_{\lambda_c^\pm} = 0$, rendering $F_c$ differentiable. Computing the second derivative at the boundaries we find
$\partial^2_{\lambda_c} F_c(\lambda_c^\pm) = 1/\partial_{j_c}^2I_c(\partial_{j_c} I_c^{-1}(\lambda_c^\pm)) > 0,$
where positivity follows from strict convexity of $I_c$. The mismatch of second derivatives shows that non-differentiability  at the origin  of a strictly convex charge rate function corresponds to a second order dynamical phase transition.

\section{Conclusion and discussion}
\label{sec13}
We have studied charge fluctuations of single-file dynamics with inert charges in bipartite initial ensembles, extending the results of \cite{Krajnik2022SF} to generic charge measures. We found an exact finite-time dressing transformation between the particle and charge fluctuations in terms of a conditional probability, easily expressible in terms of the Fourier transform of the charge measure \eqref{sum_dist}. The dressing takes on its simplest form at the level of full-counting statistics where it reduces to a simple substitution rule \eqref{Dg_rule}.
By analyzing the asymptotics of the dressing transformation, we were able to study typical and large charge fluctuations with only mild technical assumption on the underlying particle fluctuations and the charge measure.

Typical charge fluctuations depend only on the first two moment of the charge measure and are generically Gaussian \eqref{typ_charge_gauss}, except in equilibrium ensembles with vanishing mean charge, where they are parametrically suppressed and anomalous, being distributed according to an M-Wright distribution \eqref{typ_charge_wright}. The analysis of large charge fluctuations is technically more involved, incorporating asymptotic analysis of the integral representation of the conditional rate function and subsequent analysis of the convex minimization problem \eqref{Ic_minimization}. The outcome is a charge rate function that is either differentiable, has a non-convex cusp at a finite charge current, a convex non-differentiable point at the origin or a combination of the latter two. In terms of the charge scaled cumulant generating function, the cusp corresponds to first order dynamical phase transition while non-differentiability at the origin is mapped to a second order dynamical phase transition. 
Our results show that charged single-file dynamics constitute a class of dynamical systems with distinct universal fluctuation phenomenology, with Gaussian charges the exactly solvable model of the class. 

As pointed out in \cite{Krajnik2022SF,Krajnik2024,PhysRevB.109.024417}, dynamical two-point functions are not sufficient to unambiguously determine an effective evolution law. The exact dynamical charge two-point of arguably the simplest charged single-file system has been derived in \cite{Klobas2018} and is diffusive in a half-filled equilibrium ensemble, which might erroneously suggest an effective diffusion equation. This is however incompatible with anomalous fluctuations in the system \cite{Krajnik2022}. 

At the level of two-point functions a hydrodynamic description of similar kinetically constrained models has been put forward in \cite{Knap}. Similar ideas have also appeared in a hydrodynamic study of fluctuations of the large anisotropy limit of the easy-axis XXZ spin chain \cite{PhysRevB.109.024417}, which also finds anomalous spin fluctuations following an M-Wright distribution, observed numerically also at finite anisotropy \cite{Krajnik2024}. While such systems do not manifestly satisfy a single-file constraint, they can be mapped to a single-file dynamics with non-inert charges by a bond-site transformation \cite{10.21468/SciPostPhysCore.4.2.010,10.21468/SciPostPhys.10.5.099,PhysRevE.104.044106}.

An important open problem is to understand how much of the phenomenology observed for inert charges remains intact when charges influence the particle dynamics. 
Recently, a symmetry-based approach has been used to derive a three-mode hydrodynamic theory of Dirac fluids \cite{gopalakrishnan2024nongaussian} which found anomalous charge current fluctuations described by an M-Wright distribution in absence of normal diffusion.

Another question concerns the development of a hydrodynamic framework for typical fluctuations in integrable systems. While ballistic macroscopic fluctuation theory \cite{BMFT} and space-time duality \cite{SpaceTimeDuality} give access to large fluctuations, the central limit property critically hinges on regularity of the full-counting statistics \cite{Bryc1993}, which has been found to be violated in certain integrable models that exhibit sub-ballistic transport at half-filling \cite{Krajnik2022a,Krajnik2024}. 
Very recently, a set of three-mode hydrodynamic equations, akin to those describing Dirac fluids, has been derived microscopically in a deterministic single-file cellular automaton \cite{yoshimura2024} and used to show anomalous fluctuations follow from Euler hydrodynamics.
It appears plausible that sub-ballistic transport in integrable models generically signals anomalous full-counting statistics.

The study of fluctuations in strongly interacting quantum systems is becoming experimentally feasible \cite{rosenberg2023dynamics} and could provide insight into the seemingly related fluctuation phenomenology.

\paragraph*{\bf Acknowledgments}
We thank Robert Konik whose question motivated the present study, Johannes Schmidt, Vincent Pasquier, Enej Ilievski and Tomaž Prosen for collaboration on related projects and comments on the manuscript, Takato Yoshimura for discussions on regularity in generalized hydrodynamics and the anonymous referee for their comments which helped to improve the manuscript.
We gratefully acknowledges hospitality of the Simons Center for Geometry and Physics, Stony Brook University during the ``Fluctuations, Entanglements, and Chaos: Exact Results'' workshop and program in September 2023 where this work was initiated. ŽK is supported by the Simons Foundation via a Simons Junior Fellowship grant 1141511.


\appendix

	\section{Dressing the full-counting statistics}
\label{app:MGF_dressing}
Introducing the counting fields $z_i = e^{\lambda_i}$ and writing out the conjugation in Eq.~\eqref{D_D_rel} we find
\begin{equation}
	G_{c,p}(z_c, z_p|t) = \frac{1}{2\pi \ii} \ointctrclockwise_{|z|=1} \frac{\dd {z}}{{z}}K(z_c, z_p|z)G_p({z}^{-1}|t), \label{app_Gcp_int}
\end{equation}
where we noted that $G_p(z|t)$ is a convergent Lauren series in $z$ on $\mathbb{C}\backslash \{0,\infty\}$ and  interchanged the order of summation over $z^p$ and integration to arrive at the kernel $K$ of the dressing operator
\begin{equation}
	K(z_c, z_p|z) = \sum_{J_p=-\infty}^{\infty} \int_\mathbb{R} {\dd J_c}\, z_c^{J_c} \mathcal{P}_{c|p}(J_c|J_p) z_p^{J_p} z^{J_p}. \label{app:K_def}
\end{equation}
The problem in evaluating \eqref{app_Gcp_int} is the lack of convergence of \eqref{app:K_def} in the $z$-plane. We therefore introduce the projectors $\mathscr{P}^{(z)}_\pm$ onto the non-negative/positive powers in the Laurent expansion of a function $f(z)$
\begin{equation}
	\mathscr{P}_{z}^\pm[f(z)] = \frac{1}{2\pi \ii}  \sum_{n=0}^\infty z^{\pm n} \ointctrclockwise_{|z|=1} \frac{\dd z}{z} z^{\mp n} f(z)
\end{equation}
and define the projected kernels
\begin{equation}
	K^{\pm}(z_c, z_p|z) = \mathscr{P}_{z}^\pm[K(z_c, z_p|z)]. \label{K_pmdef}
\end{equation}
Introducing the corresponding projected dressed full-counting statistics
\begin{equation}
	G_{c, p}^{\pm}(z_c,z_p|t) = \frac{1}{2\pi \ii} \ointctrclockwise_{|z|=R_\pm} \frac{\dd {z}}{{z}}K^{\pm}(z_c, z_p|z^{\mp 1})G_p({z^{\pm 1}}|t), \label{G_pmdef}
\end{equation}
where $R_\pm$ denote the radia of convergence of $K^{{\pm}}(z_c, z_p|z^{\mp 1})$ respectively, the integral \eqref{app_Gcp_int} decomposes as
\begin{equation}
	1+G_{c,p}(z_c, z_p|t) = G^{+}_{c,p}(z_c, z_p|t) + G^{-}_{c,p}(z_c, z_p|t). \label{G_pmsplit}
\end{equation}
Note that we have inverted the integration variable $z \to 1/z$ in the definition of $G^{+}_{c,p}$ \eqref{G_pmdef} to ensure that the integral kernel has a non-vanishing region of convergence around $z \to \infty$. The additional one in the left-hand side of \eqref{G_pmsplit} comes from both projectors $\mathscr{P}^\pm$ picking up the constant term $G_p(z^0|t) = 1$. Inserting the Fourier form of the conditional probability given by Eq.~\eqref{sum_dist} into \eqref{K_pmdef}, interchanging the orders of summation and integration, summing up the power series and recalling that $z_c=e^{\lambda_c}$, we find
\begin{equation}
	K^{\pm}(z_c, z_p | z^{\mp 1}) = \int_{\mathbb{R}}\dd J_c\, \int_{\mathbb{R}} \frac{\dd k}{2\pi} \frac{e^{(\lambda_c  \pm \ii k) J_c}}{1 - \hat \omega_{\mu_\mp}(k)z_p^{\pm 1}/z},
\end{equation}
for $|z|>|\hat \omega_{\mu_\mp}(k)z_p^{\pm 1}|$.
Interchanging the order of integration and carrying out the integral over $J_c$ yields
\begin{equation}
	K^{\pm}(z_c, z_p | z^{\mp 1}) = \int_{\mathbb{R}} \dd k\,  \frac{\delta(\lambda_c \pm \ii k)}{1 - \hat \omega_{\mu_\mp}(k)z_p^{\pm 1}/z},
\end{equation}
for $|z|>|\hat \omega_{\mu_\mp}(k)z_p^{\pm 1}|$,
which trivially integrates to
\begin{equation}
	K^{\pm}(z_c, z_p | z^{\mp 1}) = \frac{1}{1 - z_\pm/z}, \quad {\rm for}\  |z|> |z_\pm|,
\end{equation}
where $z_\pm = \hat \omega_{\mu_\mp}(\pm \ii \lambda_c)z_p^{\pm 1}$. We now return to the integral in \eqref{G_pmdef} and evaluate it over a contour $|z| >   \max\ \{R_-, R_+ \}$ by the residue theorem. The contour encircles the two poles at $z = 0$ and $z = z_\pm$ with residues
\begin{align}
	&{\rm Res}\left(\frac{z^m}{z-z_\pm},\ z=0 \right) = 
	\begin{cases}
		0 &{\rm for}\ m \geq 0,\\
		- z_\pm^{m} &{\rm for}\ m < 0.
	\end{cases},\\
 &{\rm Res}\left(\frac{z^m}{z-z_\pm},\ z=z_\pm \right) = z_\pm^m.
\end{align}
For non-negative powers of $z^{m \geq 0}$, the two residues cancel while positive powers amount to a simple replacement $z^{m \geq 0} \mapsto z_\pm^{m}$. Recombining the split charge-particle full-counting statistics \eqref{G_pmsplit} then gives a straightforward substitution
\begin{equation}
	G_{c, p}(z_c, z_p|t) = G_{p}(z_p|t)|_{z_p^{\pm n} \mapsto z_p^{\pm n} \hat \omega^n_{\mu_\mp}(\pm \ii \lambda_c)}, \quad n \in \mathbb{N}, \label{app:substitution}
\end{equation}
where we note that the additional one in \eqref{G_pmsplit} is canceled by the constant term being dressed in both $G^{\pm}_{c,p}$ terms. The substitution \eqref{app:substitution} gives the action of the dressing operator $\mathfrak{D}_G$ acting on $G_p$, see Eq.~\eqref{Dg_rule}.

\section{Typical fluctuations}
\label{app:typ_cond}
Typical fluctuations are governed by the large-scale structure of the measure which corresponds to small $k$-behavior of the Fourier transformed measure $\hat \omega_\mu$
\begin{equation}
	\hat \omega_\mu(k) = 1 - \ii \mu k - \frac{1}{2} (\sigma^2_\omega + \mu^2) k^2 + \mathcal{O}(k^3). \label{Taylor_measure}
\end{equation}
\subsection{Nonequilibrium typical conditional probability}
\label{app:typ_cond1}
To evaluate the typical conditional probability we start with the integral representation Eq.~\eqref{sum_dist} and change variables to $u=kt^{\alpha/2}$
\begin{widetext}
	
\begin{equation}
	\mathcal{P}^{\rm typ}_{c|p}({j}_c|{j}_p) =  \int_{\mathbb{R}} \frac{\dd u}{2\pi} \lim_{t \to \infty} \left[\hat \omega_{\mu_\nu}(ut^{-\alpha/2})\right]^{-\overline \nu (s_1^{(p)} t^{\alpha} + {j}_p t^{\alpha/2})} e^{-\ii \overline \nu u (s_1^{(c)} t^{\alpha/2}+{j}_c)}. \label{app:int2}
\end{equation}

\end{widetext}
where $\overline \nu = -{\rm sgn}(j_p)$.
We now observe that the argument of the Fourier transformed measure becomes small as $t \to \infty$, suggesting the use of the Taylor expansion \eqref{Taylor_measure}. Since the limit
	\begin{equation}
	\lim_{t \to \infty} \left[1+a_1 t^{-\alpha/2} + a_2 t^{-\alpha}\right]^{b_1 t^{\alpha} + b_2 t^{\alpha/2}} e^{-c_1 t^{\alpha/2}}
	\end{equation} 
diverges for $c_1 \neq a_1 b_1$ while it equals $e^{-\frac{a_1^2 b2}{2} + a_2 b_1 + a_1 b_2}$ for $c_1 = a_1b_1$.
It follows the conditional distribution \eqref{app:int2} is finite only if the average charge and particle currents are related by Eq.~\eqref{avg_c_p}. For these values, the integral \eqref{app:int2} simplifies
\begin{align}
	\mathcal{P}^{\rm typ}_{c|p}({j}_c|{j}_p) = \frac{e^{-\frac{({j}_c-{j}_p \mu_{\overline \nu})^2}{2 \sigma_\omega^2 |s_1^{(p)}|}}}{\sqrt{2\pi \sigma_\omega^2 |s_1^{(p)}|}}. \label{app:int3}
\end{align}

\subsection{Equilibrium typical conditional probability}
\label{app:typ_cond2}
To evaluate the typical condition probability we again start with the integral representation Eq.~\eqref{sum_dist}, which, after changing variables and $u = kt^{\alpha/4}$ and inserting the Taylor expansion \eqref{Taylor_measure} becomes
\begin{equation}
	\mathcal{P}^{\rm typ, eq}_{c|p}({j}_c|{j}_p) = \int_{\mathbb{R}} \frac{\dd u}{2\pi} e^{-\ii \nu u j_c}  \lim_{t \to \infty} \left[ \hat \omega_0(ut^{-\alpha/4}) \right]^{-\nu j_p t^{\alpha/2}}
\end{equation}
Using the definition of the exponential, $\lim_{t \to \infty} \left[1 + at^{-\alpha/2} \right]^{bt^{\alpha/2}} = e^{ab}$, we find
\begin{equation}
	\mathcal{P}^{\rm typ, eq}_{c|p}({j}_c|{j}_p) =
	 \frac{e^{-j_c^2/2|j_p|\sigma_\omega^2}}{\sqrt{2\pi|j_p|\sigma_\omega^2}}.
\end{equation}

\section{Integral representation of $M_{1/4}$}
\label{app:MWright}
The integral representation of the M-Wright functions $M_{1/4}$ is obtained starting from Theorem 2 of Ref.~\cite{Mainardi_2020}
\begin{equation}
	\frac{1}{2}M_{\nu}(|x|) =  \int_{\mathbb{R}} \frac{\dd k}{2\pi} e^{-\ii k |x|} E_{2\nu}(-k^2), \label{M_E_rel}
\end{equation}
valid for $ 0  \leq \nu \leq 1$, where $E_{\nu}(z)$ is a Mittag–Leffler function \cite{MittagLeffler}, defined for $\nu >0$, $z \in \mathbb{C}$ by the series
\begin{equation}
		E_{\nu} (z) \equiv \sum_{n=0}^\infty \frac{z^n}{\Gamma(\nu n+1)}.
\end{equation} 
For special values of $\nu$, the series can be computed explicitly. In particular, we note that $E_{1/2}(-z) = e^{z^2}{\rm erfc}(z)$, see e.g.~\cite{MittagLeffler}. Using the identity $e^{z^2}{\rm erfc}(z) = \int_0^\infty \frac{\dd y}{\sqrt{\pi}}\, e^{-zy - y^2/4}$ and taking $\nu = 1/4$ in Eq.~\eqref{M_E_rel} we find
\begin{equation}
	M_{1/4}(|x|) =  \int_{\mathbb{R}} \frac{\dd k}{\pi^{3/2}} e^{-\ii k |x|} \int_0^\infty \dd y\, e^{-yk^2 -y^2/4}.
\end{equation}
Interchanging the order of integration by Fubini's theorem, carrying out the Gaussian integral over $k$ and symmetrizing the remaining integral over the real line we come to
\begin{equation}
	M_{1/4}(|x|) =  \frac{1}{2\pi}\int_\mathbb{R} \dd y\, |y|^{-1/2}\, e^{-\frac{x^2}{4|y|} -\frac{y^2}{4}}. \label{integral_representation}
\end{equation}
Note that M-Wright functions on $\mathbb{R}_+$ are normalized as $\int_0^\infty\dd x\, M_\nu(x) = 1$. Normalization of the symmetrized function in Eq.~\eqref{integral_representation} is accordingly $\int_\mathbb{R}\dd x\, M_\nu(2|x|) = 1$.

\section{Large fluctuations}
\label{app:large_cond}
The qualitative properties of large charge deviation are primarily determined by the analytic properties of the Fourier transformed charge measure $\hat\omega_\mu$. Reality of $\omega_\mu$ is manifested as a reflection symmetry of $\hat \omega_\mu$ about the imaginary axis
\begin{equation}
	\overline{\hat \omega_\mu(k)} = \hat \omega_\mu(-\overline k). \label{imag_sym}
\end{equation}
The assumed rapid decay of the measure ensure that $\hat \omega_\mu$ is an entire function. As the Fourier transform of a measure, $\hat \omega_\mu$ is a positive-definite function on $\mathbb{R}$ and its norm on the real line is upper-bound by the norm at the origin 
\begin{equation}
	|\hat \omega_\mu(k)| \leq |\hat \omega_\mu(0)| = 1 \quad {\rm for}\ k \in \mathbb{R}. \label{norm_prop}
\end{equation}
For later convenience we note that the conditional probability Eq.~\eqref{sum_dist} bounds the scaled particle current $j_p$ entering Eq.~\eqref{cond_icp_int}
\begin{align}
	&j_p \geq j_c/c_+ \enspace \textrm{for}\ j_c \geq 0 \textrm{ and } j_p \geq 0, 	\label{charge_current_bound1}\\
	&j_p \leq j_c/c_- \enspace \textrm{for}\ j_c \geq 0 \textrm{ and } j_p \leq 0,\\
	&j_p \geq j_c/c_- \enspace \textrm{for}\ j_c \leq 0 \textrm{ and } j_p \geq 0,\\
	&j_p \leq j_c/c_+ \enspace \textrm{for}\ j_c \leq 0 \textrm{ and } j_p \leq 0.
	\label{charge_current_bound}
\end{align}
We accordingly define the domains $\mathcal{J}_p^\pm$ of the scaled particle current
\begin{equation}
	\mathcal{J}_p^- \equiv \left(j_p^{\rm min},\ j_p^{-}\right],  \quad \mathcal{J}_p^+ \equiv \left[j_p^+,\ j_p^{\rm max}\right),
\end{equation}
where the inner boundaries $j_p^\pm$ depend on $j_c$ for measures with finite support
\begin{align}
	j_p^{-} &= \min\left\{j_c/c_-, j_c/c_+\right\},\\
	j_p^{+} = \max\left\{j_c/c_-, j_c/c_+\right\}. \label{inner_boundary}
\end{align}
The Legendre-Fenchel transform relates the derivative of the particle rate function $I_p$ to inverse derivatives of the scaled cumulant generating function $F_p$, $\partial_{j_p}I_p(j_p) = (\partial_{\lambda_p} F_p)^{-1}(j).$
Differentiability and strict convexity of $F_p$ on $\mathbb{R}$ then combine to give divergent derivatives of the particle rate function at the boundaries of $\mathcal{J}_p$
\begin{align}
	&\partial_{j_p} I_p(j_p) \to -\infty\ \textrm{ as }\ j \to j_p^{\rm min}, \nonumber\\
	&\partial_{j_p} I_p(j_p) \to +\infty\ \textrm{ as }\ j \to j_p^{\rm max}. \label{jp_der_div}
\end{align}

\subsection{Large deviation form of split conditional rate functions}
\label{app:LD_asymptotics}
The integral representation of split conditional rate functions given by Eq.~\eqref{cond_icp_int} simplifies to
\begin{equation}
	I_{c|p}^{\pm}(j_c|j_p) = - \lim_{t \to \infty} t^{-\alpha} \log \int_{\mathbb{R}} \frac{\dd k}{2\pi} e^{t^{\alpha} f_\pm(k)}, \label{app:split_cond_icp_int}
\end{equation}
where 
\begin{equation}
	\mp f_\pm(k) = j_p \log \hat \omega_{\mu_\pm}(k) + \ii k j_c. \label{fpm_def}
\end{equation}
The asymptotics of \eqref{app:split_cond_icp_int} can be analyzed by the saddle point method, the saddle points $k_\pm$ satisfying
\begin{equation}
	j_p\, \partial_k \omega_{\mu_\pm}(k)|_{k=k_\pm}  + \ii j_c\, \hat \omega_{\mu_\pm}(k_\pm) = 0.
	\label{app:saddle_point_eq}
\end{equation}
\subsubsection{Saddle point on the imaginary axis}
We now show that the saddle point equations \eqref{app:saddle_point_eq} each have exactly one solution on the imaginary axis.
By virtue of the assumed rapid decay of the measure, the function
\begin{equation}
	r(x) \equiv \hat\omega_\mu(\ii x) = \int_{\mathcal{C}}  \dd \omega_\mu\, e^{cx}, \label{r_def}
\end{equation}
is well-defined and is manifestly real and positive for $x \in \mathbb{R}$. Suppressing inessential indices and setting $k_\pm = \ii x$, $x \in \mathbb{R}$, the saddle point equations can be written as
\begin{equation}
	h(x) -j_c/j_p = 0, \label{imag_axis_saddle_eq}
\end{equation}
where $h(x) = r'(x)/r(x)$ and $r'(x) = \partial_x r(x)$. Computing the derivative
\begin{align}
	h'(x) &= 
	\frac{r''(x)r(x) - [r'(x)]^2}{[r(x)]^2} \nonumber\\
	&= \frac{\iint_{\mathbb{R}^2}\dd^2 \omega_\mu\left(c_1^2 - c_1c_2 \right)e^{x(c_1+c_2)}}{[r(x)]^2} \nonumber \\
	&= \frac{\iint_{\mathbb{R}^2}\dd^2 \omega_\mu\left(c_1^2+c_2^2 - 2c_1c_2 \right)e^{x(c_1+c_2)}}{2[r(x)]^2} \nonumber\\
	&= \frac{\iint_{\mathbb{R}^2}\dd^2 \omega_\mu\left(c_1-c_2 \right)^2e^{x(c_1+c_2)}}{2[r(x)]^2} > 0, \label{der_f_pos}
\end{align}
we find that $h$ is continuous and strictly monotonous, with equality in \eqref{der_f_pos} being achieved only for a trivial measure supported on a single point. We also have
\begin{align}
	h(x) &= \frac{r'(x)}{r(x)} = \frac{\int_{\mathcal{C}}  \dd \omega_\mu\, c e^{cx}}{r(x)} \nonumber \\
	&= \frac{\int_{-\infty}^0 \dd \omega_\mu\, c e^{cx} + \int_{0}^\infty \dd \omega_\mu\, c e^{cx}}{r(x)}. \label{f_split}
\end{align}
Both terms in the numerator of \eqref{f_split} are non-zero and of different sign. Moreover, they can be made arbitrarily large or small by sending $x \to \pm \infty$.  If $\omega_\mu$ is finitely supported, the integrals in \eqref{f_split} localize around the support's boundary points
\begin{equation}
	h(x) \to c_- \enspace  \textrm{as}\ x \to - \infty, \qquad h(x) \to c_+ \enspace  \textrm{as}\ x \to \infty. \label{bounded_f}
\end{equation}
Enlarging the support to $\mathbb{R}$ accordingly gives divergent asymptotic values as $|x| \to \pm \infty$ 
\begin{equation}
	h(x) \to \pm \infty \enspace  \textrm{as}\ x \to \pm \infty, \label{unbounded_f}
\end{equation}
which can be understood as taking the limit $c_\pm \to \pm \infty$.
Returning to Eq.~\eqref{imag_axis_saddle_eq}, we note that the expression on the left hand side changes sign for $x \to \pm \infty$ as a a consequence of \eqref{bounded_f} and bounds on the charge current (\ref{charge_current_bound1}-\ref{charge_current_bound}).
The same conclusion holds in the case of unbounded support via \eqref{unbounded_f}.
Since $h$ is continuous and strictly monotonic, it follows that Eq.~\eqref{imag_axis_saddle_eq} has exactly one solution $x_0 \in \mathbb{R}$.
Reinstating the notation of Eqs.~\eqref{app:saddle_point_eq}, we denote these solutions on the imaginary axis as
\begin{equation}
	k_\pm = \ii \kappa_\pm, \quad \kappa_\pm \in \mathbb{R}.
\end{equation}
Note that the locations of the saddle points depends only the ratio of the currents $\kappa_\pm = \kappa_\pm(j_c/j_p)$ via \eqref{imag_axis_saddle_eq}.

\subsubsection{Saddle point localization}
Around the saddle points $\ii \kappa_\pm$ the functions $f_\pm$ read

\begin{widetext}
	
\begin{equation}
	\mp f_\pm(\ii \kappa_\pm + z) = \mp f_\pm(\ii \kappa_\pm) + \frac{z^2}{2}j_p\frac{\hat \omega''_{\mu_{\pm}}(\ii \kappa_\pm)\hat \omega_{\mu_{\pm}}(\ii \kappa_\pm) - [\hat \omega'_{\mu_{\pm}}(\ii \kappa_\pm)]^2}{[\hat \omega_{\mu_{\pm}}(\ii \kappa_\pm)]^2} + \mathcal{O}(z^3).
\end{equation}

\end{widetext}
Importantly, the coefficients at quadratic order are strictly negative (for $j_p\neq 0$) by virtue of \eqref{r_def} and \eqref{der_f_pos} so that the saddle points are non-degenerate. Now consider the contours of steepest descent $\mathcal{I}_\pm$ across the saddle points $\ii \kappa_\pm$ that are locally parallel to the real axis 
\begin{equation}
	\mathcal{I}_\pm = \{k \in \mathbb{C}\, |\, {\rm Im}\, [f_\pm(k)] = {\rm const}, \enspace \mathcal{I}_\pm|_{\kappa_\pm} \parallel \mathbb{R}\}.
\end{equation}
Since $\hat \omega_{\mu_\pm}$ are entire, we can deform the integration contour in \eqref{app:split_cond_icp_int} without changing the value of the integrals \eqref{app:split_cond_icp_int}. We accordingly deform the path of integration to the steepest descent contours $\mathcal{I_\pm}$ across the saddle points, resulting in an integral whose exponential contribution is easily evaluated as $t \to \infty$
\begin{align}
	I_{c|p}^{\pm}(j_c|j_p) &= - \lim_{t \to \infty} t^{-\alpha} \log \int_{\mathcal{I}_\pm} \frac{\dd k}{2\pi} e^{t^{\alpha} f_\pm(k)} \label{app:split_cond_rate_result} \\
	&= - f_\pm(\ii \kappa_\pm) = \pm\left[j_p \log \hat \omega_{\mu_\pm}(\ii k_\pm) - k_\pm j_c \right]. \nonumber
\end{align}
\subsection{Generic properties of the split rate functions}
\label{app:generic_properties}
We now prove \Cref{statement1,statement2,statement3,statement4}, reproduced below for convenience.

	\noindent {\bf Claim 1}: $I_{c,p}^{\pm}(j_c=0,j_p)$ approach $I_p(j_p)$ as $j_p \to 0^\mp$.
\begin{equation}
	\lim_{j_p \to 0^\mp} I_{c|p}^{\pm}(j_c = 0, j_p) = I_p(j_p). \label{app:statement1}
\end{equation} 
	\noindent{\bf Proof:} For $j_c=0$, the saddle point equation \eqref{app:saddle_point_eq} reduces to 
	$ \partial_k \hat \omega_{\mu_\pm}(k)|_{k=\ii \kappa_\pm}= 0$, showing that $\kappa_\pm$ is independent of $j_p$. The result now follows directly from taking the limit $j_p \to 0^\mp$ in Eq.~\eqref{app:split_cond_rate_result}.\\
	
	\noindent {\bf Claim 2}: Derivatives of $I_{c,p}^{\pm}(j_c=0,j_p)$ with respect to $j_p$ are finite and bounded by
	\begin{equation}
		\mp \infty < \pm \partial_{j_p} I_{c,p}^{\pm}(j_c = 0, j_p \in \mathcal{J}_p^\mp ) \leq   \pm \partial_{j_p}I_p(j_p). \label{statement2_app}
	\end{equation}\\
	\noindent{\bf Proof:} Computing $\partial_{j_p} I_{c|p}^\pm(j_c|j_p)$ from Eq.~\eqref{app:split_cond_rate_result} and simplifying we find
	\begin{equation}
	 \partial_{j_p} I_{c,p}^\pm(j_c,j_p) = \partial_{j_p}I_p(j_p) \pm \log \hat \omega_{\mu_\pm}(\ii \kappa_\pm), \label{app:jp_der}
	\end{equation}
	where reality of $I_{c,p}^\pm$ follows from \eqref{imag_sym}. It is straightforward to show that $\hat \omega_{\mu_\pm}$ is strictly convex on the imaginary axis, $\partial_x^2 \hat \omega_{\mu_\pm}(\ii x) =  \int_{-\infty}^{\infty} \dd \omega_{\mu_\pm}\, c^2 e^{c x} > 0$ while for $j_c=0$, the saddle point equation specifies that the derivative of $\hat \omega_{\mu_\pm}$ vanishes. Since the saddle points $\ii \kappa_\pm$ are on the imaginary axis they are precisely the minima of $\hat \omega_{\mu_\pm}$ on the imaginary axis. Taking the value at the origin $\hat \omega_{\mu_\pm}(0) = 1$, we have
	\begin{equation}
		-\infty<\log \hat \omega_{\mu_\pm}(\ii \kappa_\pm) \leq \log \hat \omega_{\mu_\pm}(0)= 0, \label{app:log_bounds}
	\end{equation}
	where finiteness follows from positivity of $\hat \omega_{\mu_\pm}(\ii \kappa_\pm)$.
	Plugging \eqref{app:log_bounds} into \eqref{app:jp_der} gives the result.\\
	
	\noindent {\bf Claim 3 }: Derivatives of $I_{c,p}^{\pm}(j_c\neq 0,j_p)$ with respect to $j_p$ diverge on inner boundaries of $\mathcal{J}_p^\pm$.
	\begin{equation}
	\pm \partial_{j_p} I_{c,p}^{\pm}(j_c \neq 0| j_p \to j_p^\mp) \to  \infty. \label{app:statement3}
	\end{equation}
	\noindent {\bf Proof:} The derivatives are given by \eqref{app:jp_der} for all $j_c$. Unlike in the case $j_c=0$, the saddle points $\kappa_\pm$ are now determined by the full equation \eqref{app:saddle_point_eq} or alternatively by \eqref{imag_axis_saddle_eq} (where indices are again suppressed for simplicity). We start by considering the ratio of currents in the saddle point equation. From \eqref{inner_boundary} we have the limits
	\begin{equation}
		\lim_{j_p \to j_p^\pm} j_c/j_p = 
		\begin{cases}
			c_+ & \textrm{for}\ \pm j_c > 0,\\
			c_- & \textrm{for}\ \pm j_c < 0.
		\end{cases}
	\end{equation}
	Since the above values match the limiting values \eqref{bounded_f}, we conclude that the locations of saddle points $\kappa_\pm$ diverge
	\begin{equation}
		{\rm sgn}(j_c j_p)\, \kappa_\pm \to \infty \enspace \textrm{ as } \enspace j_p \to j_p^\pm.
	\end{equation}
	It follows that $\hat \omega_{\mu_\pm}(\ii \kappa_\pm) = \int_{-\infty}^\infty \dd \omega_{\mu_\pm}(c)\, e^{\kappa_\pm c} \to \infty$ as $|\kappa_\pm| \to \infty$, which together with \eqref{app:jp_der} gives the result.\\
	
	\noindent {\bf Claim 4}: The Hessian determinant of $I_{c,p}^\pm(j_c,j_p)$ is positive.
	\begin{equation}
	\det {\rm H}_{I_{c,p}^\pm} (j_c,j_p) > 0. \label{app:statement4}
	\end{equation}
	\noindent {\bf Proof:} We introduce $u=j_c/j_p$  and compute the partial derivatives of $I^\pm_{c,p}$ 
	\begin{align}
		&\partial^2_{j_p} I_{c,p}^\pm(j_c,j_p) =  \frac{u^2 }{|j_p|} \frac{\dd }{\dd u}\kappa_\pm(u) + \partial_{j_p}^2 I_p(j_p),\\
		&\partial^2_{j_c} I_{c,p}^\pm(j_c,j_p) =  \frac{1}{|j_p|} \frac{\dd }{\dd u}\kappa_\pm(u),\\
		&\partial_{j_c}\partial_{j_p} I_{c,p}^\pm(j_c,j_p) =   \frac{u}{|j_p|} \frac{\dd }{\dd u}\kappa_\pm(u).
		\label{app:second_derivative}
	\end{align}
	By taking a derivative of the saddle-point equations \eqref{app:saddle_point_eq} we also have
	\begin{align}
		\frac{\dd }{\dd u}\kappa_\pm(u) &= -\frac{\omega_{\mu_\pm}^2(\ii \kappa_\pm)}{\partial_k^2 \hat \omega_{\mu_\pm}(\ii \kappa_\pm)\omega_{\mu_\pm}(\ii \kappa_\pm) + \left[\partial_k \hat \omega_{\mu_\pm}(\ii \kappa_\pm)\right]^2} \nonumber \\
		&= \frac{1}{h'(\kappa_\pm)} > 0, \label{kappa_der}
	\end{align}
	where positivity follows from \eqref{der_f_pos} using the identification \eqref{r_def} and noting that a derivative with respect to $k$ gives an additional factor of  $\ii$.
	The Hessian determinant then reads
	\begin{equation}
		\det {\rm H}_{I_{c,p}^\pm} (j_c,j_p)= \frac{\partial_{j_p}^2 I_p(j_p)}{|j_p|} \frac{\dd }{\dd u}\kappa_\pm(u) > 0,
	\end{equation}
	with positivity a consequence of \eqref{kappa_der} and strict convexity of $I_p$, giving the desired result.

\bibliography{SF_spins}

\end{document}